\documentclass[aps,prl,twocolumn,showpacs,floatfix,nobibnotes,nofootinbib,superscriptaddress,amsmath,amssymb,longbibliography]{revtex4-1}

\pdfoutput=1

\usepackage[english]{babel}
\usepackage[colorlinks,urlcolor={blue}]{hyperref}
\usepackage{eurosym}
\usepackage[usenames]{color}
\usepackage{graphicx}
\usepackage{amsmath}
\usepackage{amsfonts}
\usepackage{amssymb}
\usepackage{mathrsfs}
\usepackage{bm}
\usepackage{verbatim}
\usepackage{ulem}
\usepackage{siunitx} 
\usepackage{simpler-wick} 

\newcommand{\beq}{\begin{equation}}
\newcommand{\eeq}{\end{equation}}
\newcommand{\bea}{\begin{eqnarray}}
\newcommand{\eea}{\end{eqnarray}}

\newcommand{\simge}{\hspace*{0.2em}\raisebox{0.5ex}{$>$}
     \hspace{-0.8em}\raisebox{-0.3em}{$\sim$}\hspace*{0.2em}}
\newcommand{\simle}{\hspace*{0.2em}\raisebox{0.5ex}{$<$}
     \hspace{-0.8em}\raisebox{-0.3em}{$\sim$}\hspace*{0.2em}}

\begin{document}
\title{
The importance of few-nucleon forces in chiral effective field theory}
\author{C.-J.~Yang}
\email{chieh.jen@eli-np.ro}
\affiliation{Department of Physics, Chalmers University of Technology, SE-412 96
G\"oteborg, Sweden}
\affiliation{Nuclear Physics Institute of the Czech Academy of Sciences, 
              25069 \v{R}e\v{z}, Czech Republic}
\affiliation{ELI-NP, ``Horia Hulubei" National Institute for Physics and Nuclear Engineering, 30 Reactorului Street, RO-077125, Bucharest-Magurele, Romania}
\author{A. Ekstr\"om}
\affiliation{Department of Physics, Chalmers University of Technology, SE-412 96
G\"oteborg, Sweden}
\author{C. Forss\'en}
\affiliation{Department of Physics, Chalmers University of Technology, SE-412 96
G\"oteborg, Sweden}
\author{G. Hagen}
\affiliation{Physics Division, Oak Ridge National Laboratory, Oak Ridge, Tennessee 37831,
USA}
\affiliation{Department of Physics and Astronomy, University of Tennessee, Knoxville,
Tennessee 37996, USA}
\author{%
G. Rupak}
\affiliation{Department of Physics \& Astronomy and HPC$^2$ Center for Computational Sciences, 
Mississippi State
University, Mississippi State, MS 39762, USA}
\author{%
U. van Kolck}
\affiliation{
Universit\'e Paris-Saclay, CNRS/IN2P3, IJCLab, 91405 Orsay, France
}
\affiliation{
Department of Physics, University of Arizona, 
Tucson, AZ 85721, USA}

\date{\today }

\begin{abstract}
  We study the importance of few-nucleon forces in chiral effective field theory for describing many-nucleon systems. A combinatorial argument suggests that three-nucleon forces—which are conventionally regarded as next-to-next-to-leading order—should accompany the two-nucleon force already at leading order (LO) starting with mass number $A\simeq 10-20$. We find that this promotion enables the first realistic description of the $^{16}$O ground state based on a renormalization-group-invariant LO interaction. We also performed coupled-cluster calculations of the equation of state for symmetric nuclear matter and our results indicate that LO four-nucleon forces could play a crucial role for describing heavy-mass nuclei. The enhancement mechanism we found is very general and could be important also in other many-body problems.

\end{abstract}

\maketitle
\section{Introduction}
Many-body interactions emerge naturally when the degrees
of freedom are reduced from elementary particles to composite ones. 
Many-body calculations performed today incorporate such interactions alongside the two-body interaction without additional considerations. We argue here, based on a combinatorial counting, for an
increase in the relative importance of few-body interactions with the number of constituent particles.

We substantiate this argument by performing explicit calculations in nuclear physics with 
chiral effective field theory ($\chi$EFT)~\cite{Hammer:2019poc,RevModPhys.81.1773}.
$\chi$EFT promises a framework to incorporate pion physics---long believed to be important---in 
an order-by-order improvable and renormalizable description of nuclear observables. 
The dramatic improvement in computational many-nucleon methods for the last couple of decades~\cite{Hergert:2020bxy} now allows to quantitatively study the role 
of 
few-body interactions in nuclei well beyond the alpha particle~\cite{Tews:2020hgp}.

When subjected to the organizing principle of Weinberg’s power counting (WPC)\footnote{Not being renormalizable, this is sometimes referred to 
more accurately 
as Weinberg's pragmatic proposal~\cite{Griehammer2022}}~\cite{We90,We91}, Hamiltonians based on chiral two- (NN) and three-nucleon (NNN) interactions typically describe few-nucleon systems well at sufficiently high orders~\cite{Machleidt:2011zz,Epelbaum:2012vx}, but in most cases fail to predict essential bulk properties of finite nuclei as well as a realistic equation of state (EoS) of infinite matter~\cite{overbind,radius,bay4,problem,radius1}. Chiral interactions that accurately generate empirical saturation properties often provide a less accurate description of few-nucleon data~\cite{nnlosat}. The same problem is encountered when the $\Delta(1232)$ isobar---a relatively low-lying baryon excitation---is incorporated~\cite{vanKolck:1999mw}, albeit to a lesser degree~\cite{nnlodelta,weiguang20}.

Though widely adopted, WPC is plagued by renormalizability problems~\cite{vanKolck:2020llt} starting already at leading order (LO)~\cite{Kaplan:1996xu,Beane:2001bc,Nogga:2005hy,PavonValderrama:2005uj}, and persisting at higher orders~\cite{Ya09A,Ya09B,ZE12}. Renormalization-group (RG) invariance can be achieved at LO with nonperturbative one-pion exchange restricted to low partial waves and accompanied by contact interactions that are underestimated in WPC~\cite{Kaplan:1996xu,Beane:2001bc,Nogga:2005hy,PavonValderrama:2005uj,Birse}. Subleading corrections, to be treated in the distorted-wave Born approximation (DWBA)~\cite{Birse:2007sx,Long:2007vp}, yield a reasonable description of 
NN data \cite{Valdper,Valdperb,BY,BYb,BYc,bingwei18}. This renormalization  approach, which we refer to as modified Weinberg's power counting (MWPC), provides realistic 
LO and next-to-LO (NLO) predictions of the $^3$H and $^{3,4}$He 
ground-state properties~\cite{Song:2016ale,Yang:2020pgi}. However, these RG-invariant NN interactions predict unstable 
ground states in heavier nuclei such as $^{6}$Li and $^{16}$O~\cite{Yang:2020pgi}---an unrealistic feature also 
encountered~\cite{Stetcu:2006ey,Contessi:2017rww,pionless16b} in lower-energy pionless EFT~\cite{Hammer:2019poc}. Although the slow convergence in the NN $^1S_0$ channel \cite{Birse:2010jr} can be mitigated  with a dynamical dibaryon field \cite{Bs} (which can also account for the amplitude zero~\cite{SanchezSanchez:2017tws}), the resulting energy-dependent potential makes it difficult to solve the many-nucleon Schr\"odinger equation in practice. A separable, and momentum-dependent, formulation (SEP) of the $^1S_0$ dibaryon potential \cite{Bs} unfortunately 
yields results comparable to MWPC~\cite{Yang:2020pgi}. 
It thus appears that existing RG-invariant LO interactions in $\chi$EFT are also deficient. 

A common feature of existing power-counting schemes is that few-nucleon interactions enter at subleading orders. The LO role of an NNN force in pionless EFT~\cite{Bedaque:1999ve} led the authors of Refs.~\cite{Pisa,Kievsky:2018xsl} to promote a contact NNN interaction to LO also in $\chi$EFT. However, one would like to understand the promotion or demotion of interactions either on the basis of the RG coupled to naturalness~\cite{vanKolck:2020plz} or another power-counting argument. In contrast to pionless EFT, the trinucleon system is RG invariant 
without NNN interactions in $\chi$EFT 
up to NLO with either MWPC or SEP~\cite{Song:2016ale,Yang:2020pgi}.

Conspicuously absent so far from the application of EFT to heavier nuclei is any attempt to account for factors of the mass number $A\gg 1$~\cite{yang_rev}. In this paper we put forward a combinatorial argument for promoting many-nucleon interactions to LO as $A$ increases. We 
examine the quantitative consequences of this promotion for the description of $^{16}$O, $^{40}$Ca, and the EoS for symmetric nuclear matter (SNM). Our finding---the mechanism which makes higher-body interactions more important in many-particle systems---is very general and applicable to any system where the interplay between the density and the range of the interaction is non-trivial. 

\section{Theoretical arguments}
Existing power-counting schemes rely on a perturbative expansion in the ratio $Q/M_{\rm hi}\ll 1$, where $Q$ represents low-momentum scales and $M_{\rm hi} \lesssim 1$ GeV is the $\chi$EFT breakdown scale associated with nonperturbative QCD physics. The relevant low-momentum scales include the typical momentum $p$ of a nuclear process, the pion mass $m_\pi\simeq 140$ MeV, and the pion decay constant $f_\pi\simeq\SI{93}{\mega\eV}$, which are normally assumed to be similar: $Q\sim f_\pi\sim m_\pi\sim p$. Here, for simplicity, we consider Deltaless $\chi$EFT, so formally considering the Delta-nucleon mass difference to be a high-energy scale like the nucleon mass $m_N\simeq 940$ MeV.

\begin{figure}[t]
\includegraphics[width=0.40\textwidth,clip=true]{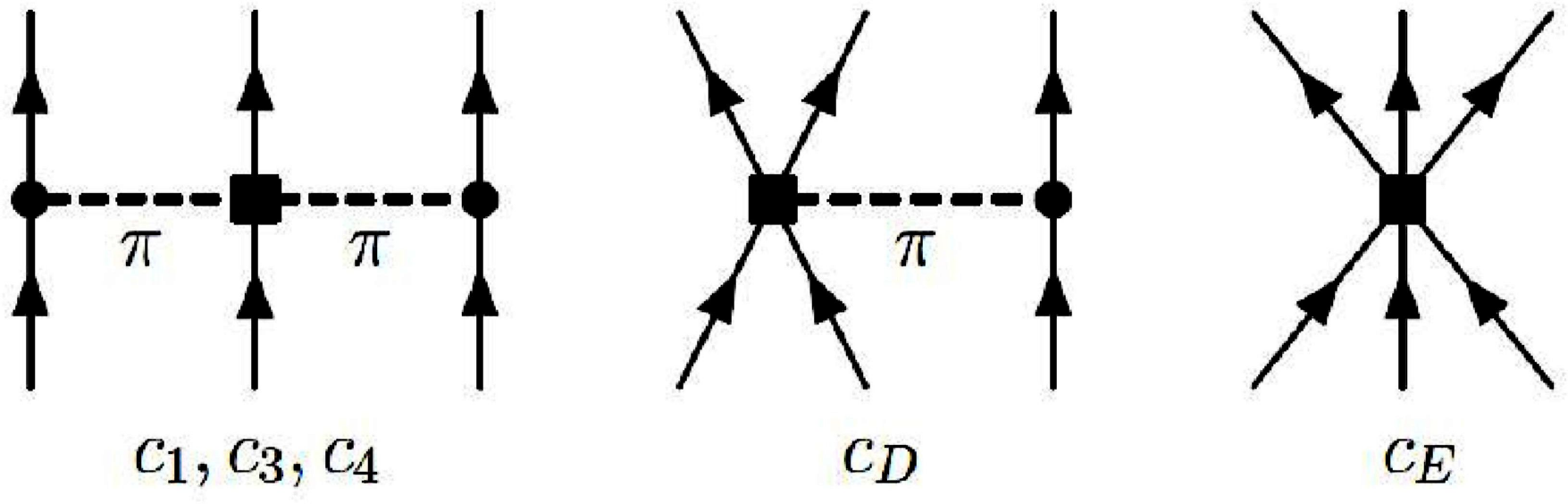}
\caption{Leading NNN interaction diagrams of pion-, pion-short, and short-range with LECs $c_{1,3,4}$, $c_D$, and $c_E$, respectively.}
\label{fig:3nf}
\end{figure}

The size of multi-nucleon interactions is usually estimated from naive dimensional analysis (NDA)~\cite{Manohar:1984,nda2,nda3,nda4}, according to which a generic 2$a$-nucleon, $p$-pion operator in the Lagrangian is
\begin{align}
    O_a&=f_\pi^2 M_\text{hi}^2\left(\frac{m_\pi}{M_\text{hi}}\right)^{2m}
    \left(\frac{\nabla}{M_\text{hi}}\right)^d
    \left(\frac{\pi}{f_\pi}\right)^p
    \left(\frac{
    N^\dagger N}
    {f_\pi^2 M_\text{hi}}\right)^a\, , \label{eq:NDA}
\end{align}
where $m$ and $d$ are non-negative integers whose values are constrained by chiral symmetry and Lorentz invariance implemented in a $Q/M_\text{hi}$ expansion. An $n$-nucleon force is constructed from combinations of $O_a$ operators with $a\leq n$. The leading NNN interactions consist of pion-, pion-short-, and short-range components \cite{vanKolck:1994yi,PhysRevC.66.064001}, as shown in Fig.~\ref{fig:3nf}. The $d=m=0$ NNN and four-nucleon (NNNN) contact forces have low-energy constants (LECs) with additional factors of $(f_\pi^2 M_\text{hi})^{2-a}$ relative to the LO NN interactions. Five- and more-nucleon contact interactions must, on spin-isospin considerations, have $d>0$, which leads to additional suppression by factors of $p/M_\text{hi}$.
%
%
At the same time, the importance of $n$-nucleon interactions can be enhanced in an $A$-nucleon system by combinatorial factors as there are more ways to construct such interactions for $2<n\lesssim A/2$.
Matrix elements of $n$-nucleon interactions multiplied by the corresponding combinatorial factor represent their total contributions in an $A$-nucleon system\footnote{For example, contributions from NN and NNN interactions appear as Eq. (A2) and (A4) of Ref. \cite{Navratil:1999pw} in {\it ab initio} calculations.}---which are the quantities one should power count.
In other words, if one confines $A$ nucleons within a finite volume where all of them interact with each other, the importance $R_n$ of the $n$-nucleon interaction relative to the NN interaction can be roughly estimated as
\begin{align}
     R_n\equiv 
     \frac{{}_A C_n}{{}_A C_2}\left(\frac{\langle N^\dagger N\rangle}{f_\pi^2 M_\text{hi}}\right)^{n-2}\, ,
\label{NDA}
\end{align}
where ${}_AC_n= A!/[n!(A-n)!]$ is the binomial coefficient and $\langle N^\dagger N\rangle$ is the single-nucleon density. Note that Eq. (\ref{NDA}) applies mainly \textit{before} saturation, where the number of interacting nucleons per volume, i.e., the nuclear density, increases with $A$. Thus, one can expect that the relative contributions between higher-body interactions and NN interactions will grow, as other counter-effects---which will be discussed later---will only weaken the combinatorial growth, but not stop it, at least before saturation. Approximating $\langle N^\dagger N\rangle$ by the saturation density $\rho_0\simeq\SI{0.16}{\femto\m^{-3}}$, the relative contribution of NNN interactions is $R_3\sim A\rho_0/(3 f_\pi^2 M_\text{hi})$. Thus one might expect these interactions to become as important as the NN force for 
$A\sim 3 f_\pi^2 M_\text{hi}/\rho_0$, which for a breakdown scale in the range $0.5\lesssim M_\text{hi}/\mathrm{GeV}\lesssim 1$ translates into a mass-number range $10\lesssim A\lesssim 20$. Likewise, the NNNN force becomes comparable to the NNN force for
$13\lesssim A\lesssim 26$.

A similar estimate, but not limited to short-range operators, results from a diagrammatic analysis, where we count pion propagators as $Q^{-2}$, nucleon propagators as $m_N Q^{-2}$, and the loop measure involving nucleons as $Q^5 (4\pi m_N)^{-1}$. In this case, the penalty for the connection to an additional nucleon is $Q/M_{\rm hi}$~\cite{Hammer:2019poc}, in agreement with the power counting of Friar \cite{Friar1997}. Combined with NDA, it leads to a suppression factor of $(Q/M_{\rm hi})^2$ instead of $\rho_0/(f_\pi^2 M_{\rm hi})$, for the leading short-range few-nucleon interactions. The two estimates are numerically consistent if $Q$ for nuclear matter is larger than $f_\pi$ by a factor $\simeq 3$. If one uses instead Weinberg's estimate \cite{We91}, where the penalty for the connection to an additional nucleon is $(Q/M_{\rm hi})^2$, the factor is instead $\simeq 4$. This argument suggests all leading three-nucleon interactions are comparable.

These estimates suffer from a number of caveats, and we expect the ratio $R_n$ to be quenched in exact calculations.
For one, they rely on NDA, which is based on purely perturbative arguments and is known to fail in the nuclear context where LO interactions must be treated nonperturbatively~\cite{vanKolck:2020plz,Griehammer2022}. For example, renormalization of LO one-pion exchange in the NN $^1S_0$ channel requires a pion-mass
dependent interaction with $a=2$, $m=1$, $d=0$, and $p=0$ \cite{Kaplan:1996xu,Beane:2001bc}, which by Eq. \eqref{eq:NDA} would appear only at next-to-NLO
(N$^2$LO). This interaction is linked by chiral symmetry to a $p=2$ interaction, and when the two pions are attached to two other nucleons it generates an enhanced NNNN force. Even with NDA, the leading NNNN interactions \cite{Epelbaum:2006eu} are not purely short ranged and are potentially more important than estimates on the basis of its short-range components. Also, detailed spin-isospin structure might reduce the appearance of certain higher-body interactions as the Pauli principle will block a subset of identical three- and four-nucleon interactions.  
Finally and probably most importantly, for large enough $A$, the enhancement of finite-range interactions is limited by the number of particles within an effective interacting volume. 
This means the growth estimated in Eq. (\ref{NDA}) will be reduced gradually 
as saturation sets in. For example, the ratio between contributions from NNN and NN interactions, as estimated from Eq. (\ref{NDA}), will approach a constant after saturation. The ratio is also modified due to the rather complicated finite-range nature of the strong interaction, its spin-isospin dependence, and the long-range Coulomb interaction.

\section{Computational evidence}
A more reliable estimate of the importance of few-nucleon interactions comes from numerical calculations. Throughout this work we start from the state-of-the-art LO and RG-invariant NN interactions MWPC40 and SEP40 constructed in Ref.~\cite{Yang:2020pgi}. The relevant LECs are fitted to reproduce the deuteron energy $E(^2\mathrm{H})= -2.22$~MeV and the $P$-wave phase shifts of the Nijmegen analysis~\cite{nij} up to laboratory energy $T_\mathrm{lab}\approx40$ MeV (center-of-mass momentum $p_\mathrm{cm}=m_{\pi}$). The additional 
$S$-wave LEC in SEP40 is fitted to reproduce the $^1S_0$ effective range $r_0=2.7$ fm. For the NNN interactions we consider the diagrams from Deltaless $\chi$EFT in Fig.~\ref{fig:3nf}. We work in momentum space and employ a non-local super-Gaussian regulator in terms of relative nucleon momenta.
The few-nucleon calculations presented in this work were carried out using the Jacobi-coordinate formulation of the no-core shell-model ~\cite{ncsm,ncsma}. The $^3$H ($^4$He) predictions were obtained in a 
harmonic-oscillator model space encompassing 41 (21) oscillator shells---that is $N_{\rm max}=40\, (20)$---and with an oscillator frequency $\hbar\omega=36$ MeV. We find that the results are convergent with respect to $N_{\rm max}$ to within 1$\%$ for regulator-cutoff values in the range $\Lambda=450-550$ MeV. 

For predicting the properties of $^{16}$O, $^{40}$Ca and SNM we employed the coupled-cluster (CC) method~\cite{kuemmel1978,cc,hagen2014}. For $^{16}$O and $^{40}$Ca our CC calculations started from a Hartree-Fock (HF) reference state expanded in a 
harmonic-oscillator basis consisting of up to 17 major shells ($N_\mathrm{max}=16$). The NNN force had an additional energy cut of $E_{3\rm max} = 16 ~\hbar\omega$, and to achieve convergent results we determined the optimal oscillator frequency for each 
model space. Furthermore, the NNN force was approximated at the normal-ordered two-body (NO2B) level which has been shown to be accurate for light- and medium-mass nuclei~\cite{normal1,normal2}. The CC calculations were performed at the $\Lambda$-CCSD(T) approximation level which includes single, double, and perturbative-triple particle-hole excitations~\cite{taube2008}. For the $\Lambda$-CCSD(T) calculations of $^{16}$O we conservatively estimate that the energies converged to within $1\%$ ($10\%$) for the regulator cutoffs $\Lambda=450,500$ (550) MeV, respectively. 
Pushing $\Lambda$ higher demands computational
resources that exceed our current capability.
Note that this limitation is of computational origin and is not due to our particular choice of power-counting scheme.
The calculations of SNM were done in the CCD(T) approximation with $A= 132$ nucleons placed in a momentum-space cubic lattice with $(2n_{\rm max}+1)^3$ mesh points for $n_{\rm max} \leq 4$ and periodic boundary conditions~\cite{cc_eos2}. Again, we approximated the NNN interaction at the NO2B level, and from calculations reported in Ref.~\cite{cc_eos2} using interactions with similar cutoffs and regulators we estimate that the effects of residual NNN interactions 
are at the order of $E/A \sim 1$~MeV for the densities considered in this work. 

To gauge the effects of chiral NNN interactions at LO in large-$A$ nuclei, we first explored leading pion-range forces governed by the $\pi$N LECs $c_{1,3,4}$. With $c_{1,3,4}=-0.74, -3.61, 2.17$ GeV$^{-1}$, as inferred from $\pi$N scattering data in Ref.~\cite{Hoferichter:2015tha}\footnote{Note that $c _{1,3,4}$ can change considerably when including higher-order corrections~\cite{Hoferichter:2015tha}.}, the net NNN contribution is repulsive in $^{16}$O, at least up to the highest cutoff (550 MeV) for which we can reliably perform CC calculations. A similar result was obtained with $c_{1,3,4}$ values from resonance saturation with the $\Delta(1232)$, which mimic the effects of the Fujita-Miyazawa force~\cite{Fujita1957} expected to be dominant in Deltaful $\chi$EFT~\cite{Hammer:2019poc}. However, a net attractive NNN force is required at LO, in MWPC, to generate a $^{16}$O 
ground state that is also stable with respect to decay into four $\alpha$ particles. We are thus led to consider also the shorter-range components of the leading NNN interactions. 

When we nonperturbatively include only the contact NNN interaction and fit the relevant LEC, $c_E$, to reproduce the triton ground-state energy $E(^3$H$)=-8.48$~MeV, we find that the $^4$He binding energy increases without any sign of convergence with respect to increasing regulator cutoff, due to the singular and attractive $c_E$ interaction at cutoffs $\simge 550$ MeV. This is in stark contrast with pionless EFT, where a contact NNN force that ensures $^3$H renormalization yields convergent results also for $^4$He~\cite{Platter:2004zs}. However, we are able to find convergent results when adding also the combined pion and short-range interaction with LEC $c_D$.

\begin{figure}[t]
\includegraphics[width=0.49\textwidth,clip=true]{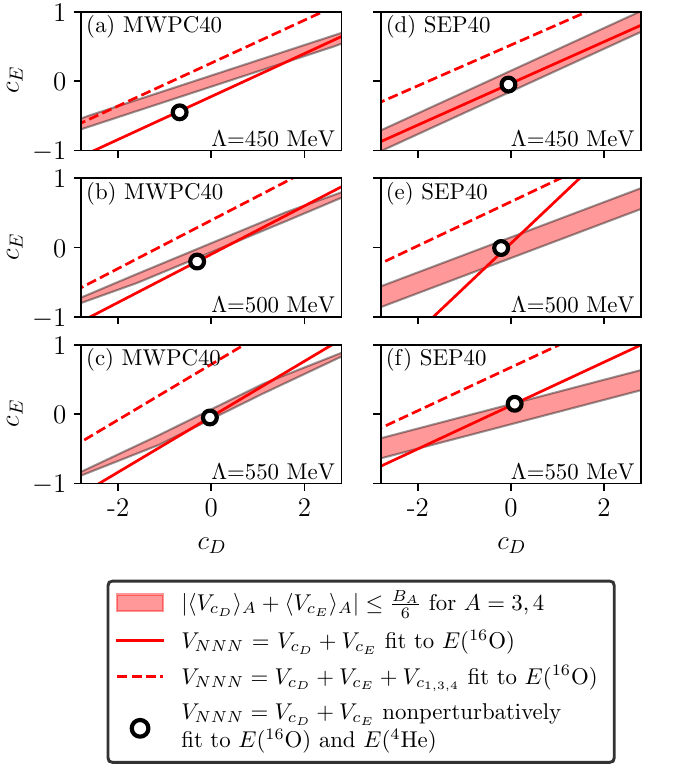}
\caption{Inferred values of the NNN LECs 
$c_{D,E}$ at various cutoff $\Lambda$ values for the 
NN potentials MWPC40 in panels (a)-(c) and SEP40 in panels (d)-(f).} 
\label{fig:cdce}
\end{figure}

It is desirable to renormalize the combination of $c_{D}$ and $c_{E}$ 
to observables of nuclei where the NNN interaction can be considered LO but NNNN interactions are not yet significant. Unfortunately, this will unavoidably involve non-trivial calculations of light-mass, open-shell nuclei.
Since estimating $c_{D,E}$ using observables obeying few-nucleon universality, such as $A=3,4$ binding energies~\cite{corr,corr1}, leads to highly degenerate solutions, we adopt instead the following procedure. First, we calculate the DWBA contributions to $^3$H and $^4$He from each of the NNN interaction terms. The perturbative treatment stems from our expectation of a small contribution from NNN interactions in few-nucleon systems. Indeed, it turns out that the net contribution from the $c_{1,3,4}$ diagrams to the binding energy is $\leq 10 \%$. We then estimate a range for $c_{D,E}$ such that the sum of their DWBA contributions, $\langle V_{c_D}\rangle_A + \langle V_{c_E}\rangle_A$, does not exceed the expected magnitude of an N$^2$LO correction according to NDA, which we conservatively estimate as $\sim 1/6$ of the corresponding binding energy $B_A$ assuming $Q/M_\text{hi} \approx 1/3$~\cite{wesolowski2021fast}. That is, we impose
\begin{eqnarray}
|\langle V_{c_D}\rangle_A +\langle V_{c_E}\rangle_A| \leq 
\frac{B_A}{6}, \quad A=3,4.
\label{res}
\end{eqnarray}
The allowed ranges are indicated as filled regions in Fig.~\ref{fig:cdce} for the two NN interactions MWPC40 and SEP40. 
Expecting a non-negligible contribution from the NNN interactions for larger $A$, we add the $c_{D,E}$ NNN interactions nonperturbatively in the $^{16}$O CC calculations. We infer a (narrow) range of values of $c_{D,E}$ values (solid lines in Fig.~\ref{fig:cdce}) for which the predicted ground-state energy falls within 10\% of the experimental value $E(^{16}$O)$\simeq -128$ MeV, only limited by a conservative CC method error and neglecting the EFT truncation error.

An overlap between the two constraints on $c_{D,E}$ exists in the cutoff range we were able to test and for values consistent with naturalness expectations. The solid lines shrink to the dots in Fig.~\ref{fig:cdce} if $c_{D,E}$ interactions are treated nonperturbatively to reproduce the experimental values of $E(^{3}$H), $E(^{4}$He) and $E(^{16}$O). Since most of the dots reside rather close to the filled regions, one could treat the $c_{D,E}$ NNN interactions nonperturbatively also in $A=3,4$ nuclei without significant consequences. When additionally promoting the pion-range NNN force it is also possible to find $c_{D,E}$ values for which $E(^{16}$O) falls within 10\% of experiment. However, the repulsive character of this NNN force shifts the $c_{D,E}$ parametrization to the dashed lines in Fig.~\ref{fig:cdce}, which do not always overlap with the range of values (filled regions) that reproduce few-nucleon binding energies.
Since Eq. (\ref{res}) is merely an estimate, we cannot rule out the inclusion of pion-range NNN forces at LO completely. Nevertheless, in the following we analyze the role of NNN interactions in many-nucleon systems using a minimal set of NNN interaction terms proportional the smallest values of $c_{D,E}$ in the overlap between the solid line and the filled area.

\begin{figure}[t]
\includegraphics[width=0.49\textwidth,clip=true]{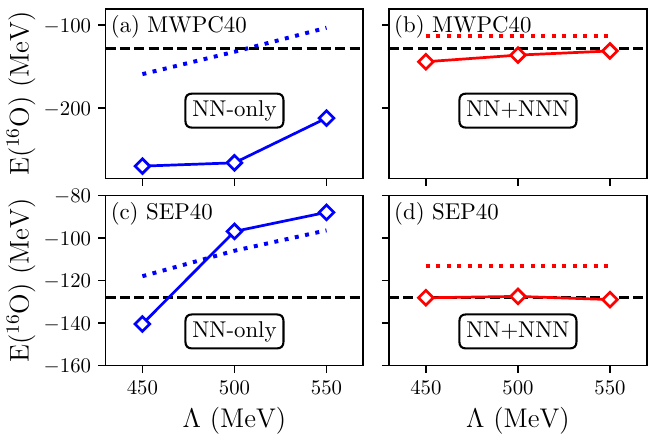}
\caption{
Ground-state energy $E$ of $^{16}$O as a function of the cutoff $\Lambda$. CC results at three different cutoffs are marked with diamonds and connected with solid lines. Results from NN-only interactions are shown in panels (a) and (c) and from NN+NNN ($c_{D,E}$) in panels (b) and (d). The NN potential MWPC40 (SEP40) is used in the top (bottom) panels.  The experimental energy (theoretical four-$\alpha$ threshold) is denoted by a dashed (dotted) line.} 
\label{fig:16O}
\end{figure}

In Fig.~\ref{fig:16O}(a),(c) we display the CC results for $E(^{16}$O) without the $c_{D,E}$ NNN interactions. The NN-only results based on MWPC40 and SEP40 at LO~\cite{Yang:2020pgi} exhibit a strong cutoff dependence, and the former interaction yields tremendous overbinding. For SEP40 and $\Lambda\simge 500$ MeV, the $^{16}$O 
ground state becomes energetically unstable with respect to decay into four $\alpha$ particles. In fact, the MWPC40 and SEP40 NN-only interactions also generate HF single-nucleon states that are starkly different from canonical shell-model expectations and allow, as we have verified numerically, for a deformed $^{16}$O 
ground state~\cite{Yang:2020pgi}.
Clearly, CC calculations including the minimal set of NNN interactions yield significantly improved results, see Fig.~\ref{fig:16O}(b),(d). Both cutoff dependence and stability with respect to four-$\alpha$ breakup are rather satisfactory throughout the examined cutoff range, especially for an EFT at LO. 
We also obtain a charge radius with a variation of about 10\% around the value $\simeq 2.1$
fm for cutoff variation in the range $\Lambda=450-550$ MeV. This is about 20\% of the experimental value and thus within the expected LO error of $\sim 30\%$.
In fact, this is the first time a realistic $^{16}$O ground state is obtained with an EFT at LO.

For the same set of NN+NNN interactions, we calculated the $^{40}$Ca ground-state energy and we obtain predictions within $15\%$ of the experimental value. For this nucleus we also estimated the CC method error $\simle 10\%$ up to $\Lambda=500$ MeV.
Although the $^{40}$Ca %
ground state is below the 10-$\alpha$ threshold, the HF single-particle spectrum implies that this state is highly deformed, a feature similar to the NN-only prediction for the $^{16}$O ground state~\cite{Yang:2020pgi}. This indicates the need for either a fine-tuning in LECs or NNNN interactions, as suggested by Eq. (\ref{NDA}). Note that N$^2$LO lattice calculations under WPC finds that an $SU(4)$-symmetric NNNN force is needed for accurate binding in $\alpha$-particle nuclei~\cite{Lahde:2013uqa}. 

To gauge the importance of NNN interactions we compare binding energies with and without them.
Table~\ref{t1} lists the binding energy per particle 
$B_A/A$ up to $^{40}$Ca. One can see that 
$B_A/A$ tends to grow linearly with $A$ with no sign of saturation if one adopts the NN-only, LO interactions. 
Clearly, NNN interactions play a crucial role to obtain a more realistic description of nuclei, albeit 
the deformed $^{40}$Ca suggests NN+NNN interactions might still be not enough to describe a realistic ground state. We emphasize that the NN+NNN results shown in Table~\ref{t1} are the outcome of the numerical calculation and subsume both the combinatorial enhancement and weakening factors such as the finite range of nuclear forces. The importance of NNN interactions for heavier nuclei is an observation that does not explicitly assume a combinatorial argument. The combinatorial argument of Eq. (\ref{NDA}) is, like other power-counting arguments, an {\it a priori} guide for the calculation. In this particular case, it offers an explanation for the observed enhancement of a certain class of otherwise subleading effects (few-nucleon interactions) beyond the alpha particle.

\begin{table}[b]
\begin{tabular}{c|cccc}
\hline
\hline
        $B_A/A$           & $^3$H & $^4$He & $^{16}$O & $^{40}$Ca  \\ \hline
 NN-only    & 3.3 & 8 & 17.5 & 31.6  \\ 
 NN+NNN     & 3.3 & 8 & 8.2 &  9.4  \\
\hline
\hline
\end{tabular}%
\caption{Binding energy per nucleon 
($B_A/A$) obtained with NN-only and NN+NNN interactions at LO. Here MWPC40 and $\Lambda$=450 MeV is adopted.}
\label{t1} 
\end{table}

\begin{figure}[t]
\includegraphics[width=0.47\textwidth,clip=true]{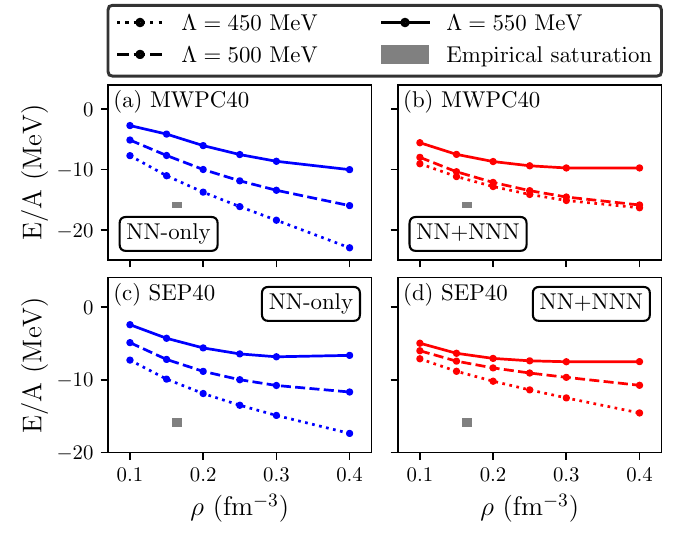}
\caption{Energy per nucleon ($E/A$) of SNM as a function of density ($\rho$) for various cutoff values $\Lambda$ and the same interactions as in Fig.~\ref{fig:16O}. The empirical saturation region~\cite{dr_eos} is marked by a grey square.}
\label{fig:EOS}
\end{figure}

The same argument implies a growing importance of NNNN interactions.
For further insight, we have calculated the EoS of SNM based on the NN-only and NN+NNN interactions devised in this work. The predictions are shown in Fig.~\ref{fig:EOS}. In most cases there is a rather strong cutoff dependence, qualitatively similar to the LO results in Ref.~\cite{Machleidt:2009bh}, and we detect no clear indication of saturation from our NN-only calculations, except with the SEP40 interaction at $\Lambda=550$ MeV (albeit far from the empirical region). Including NNN ($c_{D,E}$) interactions improves the convergence of the EoS results with $\Lambda$, however, it
does not improve agreement with empirical EoS value, at least for
$\Lambda\leq 550$ MeV.
Moreover, even if one adds a generous $2-3$ MeV uncertainty, the NN+NNN results for the EoS will still be rather far ($>5$ MeV) from the empirical saturation point. We have verified that inclusion of (repulsive) pion-range NNN interactions does not offer any improvement either.
Calculations with very high-order NN-only~\cite{Hu:2016nkw} or NN+NNN~\cite{Hebeler:2010xb,cc_eos,Drischler:2017wtt,Sammarruca:2019ncy,sam} interactions in WPC also struggle to reproduce empirical saturation properties. Unless there are substantial changes at cutoff values currently accessible only in more approximate calculations~\cite{Machleidt:2009bh}, NNNN interactions are likely to be needed at LO for describing large-$A$ nuclei, as expected from our combinatorial argument.

\section{Conclusions}
In conclusion, we found that NNN interactions are crucial for a realistic LO description of the $^{16}$O ground-state energy, and NNNN interactions are likely to be needed in 
larger-$A$ nuclei and to attain a realistic EoS for SNM. Our findings point to a missing ingredient in $\chi$EFT power counting---namely the dependence on mass number $A$ through a combinatorial enhancement of few-body interactions---that is essential for making model-independent and reliable predictions of nuclear systems. Moreover, the enhancement mechanism of many-body interactions found in this work is very general and could be important also in other many-body problems.

\textit{Acknowledgements.} 
We thank H. Grie\ss{}hammer, D. Lee, T. Papenbrock and R. Stroberg for useful discussions. This work was supported in part by the European Research Council (ERC) under the European Unions Horizon 2020 research and innovation programme (Grant agreement No. 758027); the Swedish Research Council (Grant No. 2017-04234); the Czech Science Foundation GACR grant 19-19640S and 22-14497S; the Extreme Light Infrastructure Nuclear Physics (ELI-NP) Phase II, a project co-financed by the Romanian Government and the European Union through the European Regional Development Fund - the Competitiveness Operational Programme (1/07.07.2016, COP, ID 1334);  the Romanian Ministry of Research and Innovation: PN23210105 (Phase 2, the Program Nucleu);
the U.S. Department of Energy, Office of Science, Office of Nuclear Physics, under award numbers DE-FG02-04ER41338, desc0018223 (NUCLEI SciDAC-4 collaboration), the Field Work Proposal ERKBP72 at Oak Ridge National Laboratory (ORNL); and the U.S. National Science Foundation grants PHY-1913620, PHY-2209184. The computations were enabled by resources provided by the project “eInfrastruktura CZ” (e-INFRA CZ LM2018140) supported by the Ministry of Education, Youth and Sports
of the Czech Republic, IT4Innovations at Czech National
Supercomputing Center under project number OPEN24-21 1892, CINECA under PRACE EHPC-BEN-2023B05-023, the Swedish National Infrastructure for Computing (SNIC) at C3SE and Tetralith partially funded by the Swedish Research Council.




%

%
\bibliography{3nf_ref} 

\begin{thebibliography}{82}%
\makeatletter
\providecommand \@ifxundefined [1]{%
 \@ifx{#1\undefined}
}%
\providecommand \@ifnum [1]{%
 \ifnum #1\expandafter \@firstoftwo
 \else \expandafter \@secondoftwo
 \fi
}%
\providecommand \@ifx [1]{%
 \ifx #1\expandafter \@firstoftwo
 \else \expandafter \@secondoftwo
 \fi
}%
\providecommand \natexlab [1]{#1}%
\providecommand \enquote  [1]{``#1''}%
\providecommand \bibnamefont  [1]{#1}%
\providecommand \bibfnamefont [1]{#1}%
\providecommand \citenamefont [1]{#1}%
\providecommand \href@noop [0]{\@secondoftwo}%
\providecommand \href [0]{\begingroup \@sanitize@url \@href}%
\providecommand \@href[1]{\@@startlink{#1}\@@href}%
\providecommand \@@href[1]{\endgroup#1\@@endlink}%
\providecommand \@sanitize@url [0]{\catcode `\\12\catcode `\$12\catcode
  `\&12\catcode `\#12\catcode `\^12\catcode `\_12\catcode `\%12\relax}%
\providecommand \@@startlink[1]{}%
\providecommand \@@endlink[0]{}%
\providecommand \url  [0]{\begingroup\@sanitize@url \@url }%
\providecommand \@url [1]{\endgroup\@href {#1}{\urlprefix }}%
\providecommand \urlprefix  [0]{URL }%
\providecommand \Eprint [0]{\href }%
\providecommand \doibase [0]{http://dx.doi.org/}%
\providecommand \selectlanguage [0]{\@gobble}%
\providecommand \bibinfo  [0]{\@secondoftwo}%
\providecommand \bibfield  [0]{\@secondoftwo}%
\providecommand \translation [1]{[#1]}%
\providecommand \BibitemOpen [0]{}%
\providecommand \bibitemStop [0]{}%
\providecommand \bibitemNoStop [0]{.\EOS\space}%
\providecommand \EOS [0]{\spacefactor3000\relax}%
\providecommand \BibitemShut  [1]{\csname bibitem#1\endcsname}%
\let\auto@bib@innerbib\@empty
\bibitem [{\citenamefont {Hammer}\ \emph {et~al.}(2020)\citenamefont {Hammer},
  \citenamefont {K\"onig},\ and\ \citenamefont {van Kolck}}]{Hammer:2019poc}%
  \BibitemOpen
  \bibfield  {author} {\bibinfo {author} {\bibfnamefont {H.-W.}\ \bibnamefont
  {Hammer}}, \bibinfo {author} {\bibfnamefont {S.}~\bibnamefont {K\"onig}}, \
  and\ \bibinfo {author} {\bibfnamefont {U.}~\bibnamefont {van Kolck}},\ }\href
  {\doibase 10.1103/RevModPhys.92.025004} {\bibfield  {journal} {\bibinfo
  {journal} {Rev. Mod. Phys.}\ }\textbf {\bibinfo {volume} {92}},\ \bibinfo
  {pages} {025004} (\bibinfo {year} {2020})},\ \Eprint
  {http://arxiv.org/abs/1906.12122} {arXiv:1906.12122 [nucl-th]} \BibitemShut
  {NoStop}%
\bibitem [{\citenamefont {Epelbaum}\ \emph {et~al.}(2009)\citenamefont
  {Epelbaum}, \citenamefont {Hammer},\ and\ \citenamefont
  {Mei{\ss}ner}}]{RevModPhys.81.1773}%
  \BibitemOpen
  \bibfield  {author} {\bibinfo {author} {\bibfnamefont {E.}~\bibnamefont
  {Epelbaum}}, \bibinfo {author} {\bibfnamefont {H.-W.}\ \bibnamefont
  {Hammer}}, \ and\ \bibinfo {author} {\bibfnamefont {U.-G.}\ \bibnamefont
  {Mei{\ss}ner}},\ }\href {\doibase 10.1103/RevModPhys.81.1773} {\bibfield
  {journal} {\bibinfo  {journal} {Rev. Mod. Phys.}\ }\textbf {\bibinfo {volume}
  {81}},\ \bibinfo {pages} {1773} (\bibinfo {year} {2009})},\ \Eprint
  {http://arxiv.org/abs/0811.1338} {arXiv:0811.1338 [nucl-th]} \BibitemShut
  {NoStop}%
\bibitem [{\citenamefont {Hergert}(2020)}]{Hergert:2020bxy}%
  \BibitemOpen
  \bibfield  {author} {\bibinfo {author} {\bibfnamefont {H.}~\bibnamefont
  {Hergert}},\ }\href {\doibase 10.3389/fphy.2020.00379} {\bibfield  {journal}
  {\bibinfo  {journal} {Front. in Phys.}\ }\textbf {\bibinfo {volume} {8}},\
  \bibinfo {pages} {379} (\bibinfo {year} {2020})},\ \Eprint
  {http://arxiv.org/abs/2008.05061} {arXiv:2008.05061 [nucl-th]} \BibitemShut
  {NoStop}%
\bibitem [{\citenamefont {Tews}\ \emph {et~al.}(2020)\citenamefont {Tews},
  \citenamefont {Davoudi}, \citenamefont {Ekstr\"om}, \citenamefont {Holt},\
  and\ \citenamefont {Lynn}}]{Tews:2020hgp}%
  \BibitemOpen
  \bibfield  {author} {\bibinfo {author} {\bibfnamefont {I.}~\bibnamefont
  {Tews}}, \bibinfo {author} {\bibfnamefont {Z.}~\bibnamefont {Davoudi}},
  \bibinfo {author} {\bibfnamefont {A.}~\bibnamefont {Ekstr\"om}}, \bibinfo
  {author} {\bibfnamefont {J.~D.}\ \bibnamefont {Holt}}, \ and\ \bibinfo
  {author} {\bibfnamefont {J.~E.}\ \bibnamefont {Lynn}},\ }\href {\doibase
  10.1088/1361-6471/ab9079} {\bibfield  {journal} {\bibinfo  {journal} {J.
  Phys. G}\ }\textbf {\bibinfo {volume} {47}},\ \bibinfo {pages} {103001}
  (\bibinfo {year} {2020})},\ \Eprint {http://arxiv.org/abs/2001.03334}
  {arXiv:2001.03334 [nucl-th]} \BibitemShut {NoStop}%
\bibitem [{\citenamefont {Grie{\ss}hammer}(2022)}]{Griehammer2022}%
  \BibitemOpen
  \bibfield  {author} {\bibinfo {author} {\bibfnamefont {H.~W.}\ \bibnamefont
  {Grie{\ss}hammer}},\ }\href {\doibase 10.1007/s00601-022-01739-z} {\bibfield
  {journal} {\bibinfo  {journal} {Few Body Syst.}\ }\textbf {\bibinfo {volume}
  {63}},\ \bibinfo {pages} {44} (\bibinfo {year} {2022})},\ \Eprint
  {http://arxiv.org/abs/2111.00930} {arXiv:2111.00930 [nucl-th]} \BibitemShut
  {NoStop}%
\bibitem [{\citenamefont {Weinberg}(1990)}]{We90}%
  \BibitemOpen
  \bibfield  {author} {\bibinfo {author} {\bibfnamefont {S.}~\bibnamefont
  {Weinberg}},\ }\href {\doibase 10.1016/0370-2693(90)90938-3} {\bibfield
  {journal} {\bibinfo  {journal} {Phys. Lett. B}\ }\textbf {\bibinfo {volume}
  {251}},\ \bibinfo {pages} {288} (\bibinfo {year} {1990})}\BibitemShut
  {NoStop}%
\bibitem [{\citenamefont {Weinberg}(1991)}]{We91}%
  \BibitemOpen
  \bibfield  {author} {\bibinfo {author} {\bibfnamefont {S.}~\bibnamefont
  {Weinberg}},\ }\href {\doibase 10.1016/0550-3213(91)90231-l} {\bibfield
  {journal} {\bibinfo  {journal} {Nucl. Phys. B}\ }\textbf {\bibinfo {volume}
  {363}},\ \bibinfo {pages} {3} (\bibinfo {year} {1991})}\BibitemShut {NoStop}%
\bibitem [{\citenamefont {Machleidt}\ and\ \citenamefont
  {Entem}(2011)}]{Machleidt:2011zz}%
  \BibitemOpen
  \bibfield  {author} {\bibinfo {author} {\bibfnamefont {R.}~\bibnamefont
  {Machleidt}}\ and\ \bibinfo {author} {\bibfnamefont {D.~R.}\ \bibnamefont
  {Entem}},\ }\href {\doibase 10.1016/j.physrep.2011.02.001} {\bibfield
  {journal} {\bibinfo  {journal} {Phys. Rept.}\ }\textbf {\bibinfo {volume}
  {503}},\ \bibinfo {pages} {1} (\bibinfo {year} {2011})},\ \Eprint
  {http://arxiv.org/abs/1105.2919} {arXiv:1105.2919 [nucl-th]} \BibitemShut
  {NoStop}%
\bibitem [{\citenamefont {Epelbaum}\ and\ \citenamefont
  {Mei{\ss}ner}(2012)}]{Epelbaum:2012vx}%
  \BibitemOpen
  \bibfield  {author} {\bibinfo {author} {\bibfnamefont {E.}~\bibnamefont
  {Epelbaum}}\ and\ \bibinfo {author} {\bibfnamefont {U.-G.}\ \bibnamefont
  {Mei{\ss}ner}},\ }\href {\doibase 10.1146/annurev-nucl-102010-130056}
  {\bibfield  {journal} {\bibinfo  {journal} {Ann. Rev. Nucl. Part. Sci.}\
  }\textbf {\bibinfo {volume} {62}},\ \bibinfo {pages} {159} (\bibinfo {year}
  {2012})},\ \Eprint {http://arxiv.org/abs/1201.2136} {arXiv:1201.2136
  [nucl-th]} \BibitemShut {NoStop}%
\bibitem [{\citenamefont {Binder}\ \emph {et~al.}(2014)\citenamefont {Binder},
  \citenamefont {Langhammer}, \citenamefont {Calci},\ and\ \citenamefont
  {Roth}}]{overbind}%
  \BibitemOpen
  \bibfield  {author} {\bibinfo {author} {\bibfnamefont {S.}~\bibnamefont
  {Binder}}, \bibinfo {author} {\bibfnamefont {J.}~\bibnamefont {Langhammer}},
  \bibinfo {author} {\bibfnamefont {A.}~\bibnamefont {Calci}}, \ and\ \bibinfo
  {author} {\bibfnamefont {R.}~\bibnamefont {Roth}},\ }\href {\doibase
  10.1016/j.physletb.2014.07.010} {\bibfield  {journal} {\bibinfo  {journal}
  {Phys. Lett. B}\ }\textbf {\bibinfo {volume} {736}},\ \bibinfo {pages} {119}
  (\bibinfo {year} {2014})}\BibitemShut {NoStop}%
\bibitem [{\citenamefont {Lapoux}\ \emph {et~al.}(2016)\citenamefont {Lapoux},
  \citenamefont {Som\`a}, \citenamefont {Barbieri}, \citenamefont {Hergert},
  \citenamefont {Holt},\ and\ \citenamefont {Stroberg}}]{radius}%
  \BibitemOpen
  \bibfield  {author} {\bibinfo {author} {\bibfnamefont {V.}~\bibnamefont
  {Lapoux}}, \bibinfo {author} {\bibfnamefont {V.}~\bibnamefont {Som\`a}},
  \bibinfo {author} {\bibfnamefont {C.}~\bibnamefont {Barbieri}}, \bibinfo
  {author} {\bibfnamefont {H.}~\bibnamefont {Hergert}}, \bibinfo {author}
  {\bibfnamefont {J.~D.}\ \bibnamefont {Holt}}, \ and\ \bibinfo {author}
  {\bibfnamefont {S.~R.}\ \bibnamefont {Stroberg}},\ }\href {\doibase
  10.1103/PhysRevLett.117.052501} {\bibfield  {journal} {\bibinfo  {journal}
  {Phys. Rev. Lett.}\ }\textbf {\bibinfo {volume} {117}},\ \bibinfo {pages}
  {052501} (\bibinfo {year} {2016})},\ \Eprint
  {http://arxiv.org/abs/1605.07885} {arXiv:1605.07885 [nucl-ex]} \BibitemShut
  {NoStop}%
\bibitem [{\citenamefont {Carlsson}\ \emph {et~al.}(2016)\citenamefont
  {Carlsson}, \citenamefont {Ekstr\"om}, \citenamefont {Forss\'en},
  \citenamefont {Str\"omberg}, \citenamefont {Jansen}, \citenamefont {Lilja},
  \citenamefont {Lindby}, \citenamefont {Mattsson},\ and\ \citenamefont
  {Wendt}}]{bay4}%
  \BibitemOpen
  \bibfield  {author} {\bibinfo {author} {\bibfnamefont {B.~D.}\ \bibnamefont
  {Carlsson}}, \bibinfo {author} {\bibfnamefont {A.}~\bibnamefont {Ekstr\"om}},
  \bibinfo {author} {\bibfnamefont {C.}~\bibnamefont {Forss\'en}}, \bibinfo
  {author} {\bibfnamefont {D.~F.}\ \bibnamefont {Str\"omberg}}, \bibinfo
  {author} {\bibfnamefont {G.~R.}\ \bibnamefont {Jansen}}, \bibinfo {author}
  {\bibfnamefont {O.}~\bibnamefont {Lilja}}, \bibinfo {author} {\bibfnamefont
  {M.}~\bibnamefont {Lindby}}, \bibinfo {author} {\bibfnamefont {B.~A.}\
  \bibnamefont {Mattsson}}, \ and\ \bibinfo {author} {\bibfnamefont {K.~A.}\
  \bibnamefont {Wendt}},\ }\href {\doibase 10.1103/PhysRevX.6.011019}
  {\bibfield  {journal} {\bibinfo  {journal} {Phys. Rev. X}\ }\textbf {\bibinfo
  {volume} {6}},\ \bibinfo {pages} {011019} (\bibinfo {year}
  {2016})}\BibitemShut {NoStop}%
\bibitem [{\citenamefont {Machleidt}(2018)}]{problem}%
  \BibitemOpen
  \bibfield  {author} {\bibinfo {author} {\bibfnamefont {R.}~\bibnamefont
  {Machleidt}},\ }\href@noop {} {\bibfield  {journal} {\bibinfo  {journal}
  {Nucl. Theor.}\ }\textbf {\bibinfo {volume} {37}},\ \bibinfo {pages} {62}
  (\bibinfo {year} {2018})},\ \Eprint {http://arxiv.org/abs/1901.01473}
  {arXiv:1901.01473 [nucl-th]} \BibitemShut {NoStop}%
\bibitem [{\citenamefont {Som\`a}\ \emph {et~al.}(2020)\citenamefont {Som\`a},
  \citenamefont {Navr\'atil}, \citenamefont {Raimondi}, \citenamefont
  {Barbieri},\ and\ \citenamefont {Duguet}}]{radius1}%
  \BibitemOpen
  \bibfield  {author} {\bibinfo {author} {\bibfnamefont {V.}~\bibnamefont
  {Som\`a}}, \bibinfo {author} {\bibfnamefont {P.}~\bibnamefont {Navr\'atil}},
  \bibinfo {author} {\bibfnamefont {F.}~\bibnamefont {Raimondi}}, \bibinfo
  {author} {\bibfnamefont {C.}~\bibnamefont {Barbieri}}, \ and\ \bibinfo
  {author} {\bibfnamefont {T.}~\bibnamefont {Duguet}},\ }\href {\doibase
  10.1103/PhysRevC.101.014318} {\bibfield  {journal} {\bibinfo  {journal}
  {Phys. Rev. C}\ }\textbf {\bibinfo {volume} {101}},\ \bibinfo {pages}
  {014318} (\bibinfo {year} {2020})},\ \Eprint
  {http://arxiv.org/abs/1907.09790} {arXiv:1907.09790 [nucl-th]} \BibitemShut
  {NoStop}%
\bibitem [{\citenamefont {Ekstr\"om}\ \emph {et~al.}(2015)\citenamefont
  {Ekstr\"om}, \citenamefont {Jansen}, \citenamefont {Wendt}, \citenamefont
  {Hagen}, \citenamefont {Papenbrock}, \citenamefont {Carlsson}, \citenamefont
  {Forss\'en}, \citenamefont {Hjorth-Jensen}, \citenamefont {Navr\'atil},\ and\
  \citenamefont {Nazarewicz}}]{nnlosat}%
  \BibitemOpen
  \bibfield  {author} {\bibinfo {author} {\bibfnamefont {A.}~\bibnamefont
  {Ekstr\"om}}, \bibinfo {author} {\bibfnamefont {G.~R.}\ \bibnamefont
  {Jansen}}, \bibinfo {author} {\bibfnamefont {K.~A.}\ \bibnamefont {Wendt}},
  \bibinfo {author} {\bibfnamefont {G.}~\bibnamefont {Hagen}}, \bibinfo
  {author} {\bibfnamefont {T.}~\bibnamefont {Papenbrock}}, \bibinfo {author}
  {\bibfnamefont {B.~D.}\ \bibnamefont {Carlsson}}, \bibinfo {author}
  {\bibfnamefont {C.}~\bibnamefont {Forss\'en}}, \bibinfo {author}
  {\bibfnamefont {M.}~\bibnamefont {Hjorth-Jensen}}, \bibinfo {author}
  {\bibfnamefont {P.}~\bibnamefont {Navr\'atil}}, \ and\ \bibinfo {author}
  {\bibfnamefont {W.}~\bibnamefont {Nazarewicz}},\ }\href {\doibase
  10.1103/PhysRevC.91.051301} {\bibfield  {journal} {\bibinfo  {journal} {Phys.
  Rev. C}\ }\textbf {\bibinfo {volume} {91}},\ \bibinfo {pages} {051301}
  (\bibinfo {year} {2015})},\ \Eprint {http://arxiv.org/abs/1502.04682}
  {arXiv:1502.04682 [nucl-th]} \BibitemShut {NoStop}%
\bibitem [{\citenamefont {van Kolck}(1999)}]{vanKolck:1999mw}%
  \BibitemOpen
  \bibfield  {author} {\bibinfo {author} {\bibfnamefont {U.}~\bibnamefont {van
  Kolck}},\ }\href {\doibase 10.1016/S0146-6410(99)00097-6} {\bibfield
  {journal} {\bibinfo  {journal} {Prog. Part. Nucl. Phys.}\ }\textbf {\bibinfo
  {volume} {43}},\ \bibinfo {pages} {337} (\bibinfo {year} {1999})},\ \Eprint
  {http://arxiv.org/abs/nucl-th/9902015} {arXiv:nucl-th/9902015} \BibitemShut
  {NoStop}%
\bibitem [{\citenamefont {Ekstr\"om}\ \emph {et~al.}(2018)\citenamefont
  {Ekstr\"om}, \citenamefont {Hagen}, \citenamefont {Morris}, \citenamefont
  {Papenbrock},\ and\ \citenamefont {Schwartz}}]{nnlodelta}%
  \BibitemOpen
  \bibfield  {author} {\bibinfo {author} {\bibfnamefont {A.}~\bibnamefont
  {Ekstr\"om}}, \bibinfo {author} {\bibfnamefont {G.}~\bibnamefont {Hagen}},
  \bibinfo {author} {\bibfnamefont {T.~D.}\ \bibnamefont {Morris}}, \bibinfo
  {author} {\bibfnamefont {T.}~\bibnamefont {Papenbrock}}, \ and\ \bibinfo
  {author} {\bibfnamefont {P.~D.}\ \bibnamefont {Schwartz}},\ }\href {\doibase
  10.1103/PhysRevC.97.024332} {\bibfield  {journal} {\bibinfo  {journal} {Phys.
  Rev. C}\ }\textbf {\bibinfo {volume} {97}},\ \bibinfo {pages} {024332}
  (\bibinfo {year} {2018})},\ \Eprint {http://arxiv.org/abs/1707.09028}
  {arXiv:1707.09028 [nucl-th]} \BibitemShut {NoStop}%
\bibitem [{\citenamefont {Jiang}\ \emph {et~al.}(2020)\citenamefont {Jiang},
  \citenamefont {Ekstr\"om}, \citenamefont {Forss\'en}, \citenamefont {Hagen},
  \citenamefont {Jansen},\ and\ \citenamefont {Papenbrock}}]{weiguang20}%
  \BibitemOpen
  \bibfield  {author} {\bibinfo {author} {\bibfnamefont {W.~G.}\ \bibnamefont
  {Jiang}}, \bibinfo {author} {\bibfnamefont {A.}~\bibnamefont {Ekstr\"om}},
  \bibinfo {author} {\bibfnamefont {C.}~\bibnamefont {Forss\'en}}, \bibinfo
  {author} {\bibfnamefont {G.}~\bibnamefont {Hagen}}, \bibinfo {author}
  {\bibfnamefont {G.~R.}\ \bibnamefont {Jansen}}, \ and\ \bibinfo {author}
  {\bibfnamefont {T.}~\bibnamefont {Papenbrock}},\ }\href {\doibase
  10.1103/PhysRevC.102.054301} {\bibfield  {journal} {\bibinfo  {journal}
  {Phys. Rev. C}\ }\textbf {\bibinfo {volume} {102}},\ \bibinfo {pages}
  {054301} (\bibinfo {year} {2020})},\ \Eprint
  {http://arxiv.org/abs/2006.16774} {arXiv:2006.16774 [nucl-th]} \BibitemShut
  {NoStop}%
\bibitem [{\citenamefont {van Kolck}(2020{\natexlab{a}})}]{vanKolck:2020llt}%
  \BibitemOpen
  \bibfield  {author} {\bibinfo {author} {\bibfnamefont {U.}~\bibnamefont {van
  Kolck}},\ }\href {\doibase 10.3389/fphy.2020.00079} {\bibfield  {journal}
  {\bibinfo  {journal} {Front. in Phys.}\ }\textbf {\bibinfo {volume} {8}},\
  \bibinfo {pages} {79} (\bibinfo {year} {2020}{\natexlab{a}})},\ \Eprint
  {http://arxiv.org/abs/2003.06721} {arXiv:2003.06721 [nucl-th]} \BibitemShut
  {NoStop}%
\bibitem [{\citenamefont {Kaplan}\ \emph {et~al.}(1996)\citenamefont {Kaplan},
  \citenamefont {Savage},\ and\ \citenamefont {Wise}}]{Kaplan:1996xu}%
  \BibitemOpen
  \bibfield  {author} {\bibinfo {author} {\bibfnamefont {D.~B.}\ \bibnamefont
  {Kaplan}}, \bibinfo {author} {\bibfnamefont {M.~J.}\ \bibnamefont {Savage}},
  \ and\ \bibinfo {author} {\bibfnamefont {M.~B.}\ \bibnamefont {Wise}},\
  }\href {\doibase 10.1016/0550-3213(96)00357-4} {\bibfield  {journal}
  {\bibinfo  {journal} {Nucl. Phys. B}\ }\textbf {\bibinfo {volume} {478}},\
  \bibinfo {pages} {629} (\bibinfo {year} {1996})},\ \Eprint
  {http://arxiv.org/abs/nucl-th/9605002} {arXiv:nucl-th/9605002} \BibitemShut
  {NoStop}%
\bibitem [{\citenamefont {Beane}\ \emph {et~al.}(2002)\citenamefont {Beane},
  \citenamefont {Bedaque}, \citenamefont {Savage},\ and\ \citenamefont {van
  Kolck}}]{Beane:2001bc}%
  \BibitemOpen
  \bibfield  {author} {\bibinfo {author} {\bibfnamefont {S.~R.}\ \bibnamefont
  {Beane}}, \bibinfo {author} {\bibfnamefont {P.~F.}\ \bibnamefont {Bedaque}},
  \bibinfo {author} {\bibfnamefont {M.~J.}\ \bibnamefont {Savage}}, \ and\
  \bibinfo {author} {\bibfnamefont {U.}~\bibnamefont {van Kolck}},\ }\href
  {\doibase 10.1016/S0375-9474(01)01324-0} {\bibfield  {journal} {\bibinfo
  {journal} {Nucl. Phys. A}\ }\textbf {\bibinfo {volume} {700}},\ \bibinfo
  {pages} {377} (\bibinfo {year} {2002})},\ \Eprint
  {http://arxiv.org/abs/nucl-th/0104030} {arXiv:nucl-th/0104030} \BibitemShut
  {NoStop}%
\bibitem [{\citenamefont {Nogga}\ \emph {et~al.}(2005)\citenamefont {Nogga},
  \citenamefont {Timmermans},\ and\ \citenamefont {van Kolck}}]{Nogga:2005hy}%
  \BibitemOpen
  \bibfield  {author} {\bibinfo {author} {\bibfnamefont {A.}~\bibnamefont
  {Nogga}}, \bibinfo {author} {\bibfnamefont {R.~G.~E.}\ \bibnamefont
  {Timmermans}}, \ and\ \bibinfo {author} {\bibfnamefont {U.}~\bibnamefont {van
  Kolck}},\ }\href {\doibase 10.1103/PhysRevC.72.054006} {\bibfield  {journal}
  {\bibinfo  {journal} {Phys. Rev. C}\ }\textbf {\bibinfo {volume} {72}},\
  \bibinfo {pages} {054006} (\bibinfo {year} {2005})},\ \Eprint
  {http://arxiv.org/abs/nucl-th/0506005} {arXiv:nucl-th/0506005} \BibitemShut
  {NoStop}%
\bibitem [{\citenamefont {Pav\'on~Valderrama}\ and\ \citenamefont
  {Ruiz~Arriola}(2006)}]{PavonValderrama:2005uj}%
  \BibitemOpen
  \bibfield  {author} {\bibinfo {author} {\bibfnamefont {M.}~\bibnamefont
  {Pav\'on~Valderrama}}\ and\ \bibinfo {author} {\bibfnamefont
  {E.}~\bibnamefont {Ruiz~Arriola}},\ }\href {\doibase
  10.1103/PhysRevC.74.064004} {\bibfield  {journal} {\bibinfo  {journal} {Phys.
  Rev. C}\ }\textbf {\bibinfo {volume} {74}},\ \bibinfo {pages} {064004}
  (\bibinfo {year} {2006})},\ \bibinfo {note} {[Erratum: Phys.Rev.C 75, 059905
  (2007)]},\ \Eprint {http://arxiv.org/abs/nucl-th/0507075}
  {arXiv:nucl-th/0507075} \BibitemShut {NoStop}%
\bibitem [{\citenamefont {Yang}\ \emph
  {et~al.}(2009{\natexlab{a}})\citenamefont {Yang}, \citenamefont {Elster},\
  and\ \citenamefont {Phillips}}]{Ya09A}%
  \BibitemOpen
  \bibfield  {author} {\bibinfo {author} {\bibfnamefont {C.~J.}\ \bibnamefont
  {Yang}}, \bibinfo {author} {\bibfnamefont {C.}~\bibnamefont {Elster}}, \ and\
  \bibinfo {author} {\bibfnamefont {D.~R.}\ \bibnamefont {Phillips}},\ }\href
  {\doibase 10.1103/PhysRevC.80.034002} {\bibfield  {journal} {\bibinfo
  {journal} {Phys. Rev. C}\ }\textbf {\bibinfo {volume} {80}},\ \bibinfo
  {pages} {034002} (\bibinfo {year} {2009}{\natexlab{a}})},\ \Eprint
  {http://arxiv.org/abs/0901.2663} {arXiv:0901.2663 [nucl-th]} \BibitemShut
  {NoStop}%
\bibitem [{\citenamefont {Yang}\ \emph
  {et~al.}(2009{\natexlab{b}})\citenamefont {Yang}, \citenamefont {Elster},\
  and\ \citenamefont {Phillips}}]{Ya09B}%
  \BibitemOpen
  \bibfield  {author} {\bibinfo {author} {\bibfnamefont {C.~J.}\ \bibnamefont
  {Yang}}, \bibinfo {author} {\bibfnamefont {C.}~\bibnamefont {Elster}}, \ and\
  \bibinfo {author} {\bibfnamefont {D.~R.}\ \bibnamefont {Phillips}},\ }\href
  {\doibase 10.1103/PhysRevC.80.044002} {\bibfield  {journal} {\bibinfo
  {journal} {Phys. Rev. C}\ }\textbf {\bibinfo {volume} {80}},\ \bibinfo
  {pages} {044002} (\bibinfo {year} {2009}{\natexlab{b}})},\ \Eprint
  {http://arxiv.org/abs/0905.4943} {arXiv:0905.4943 [nucl-th]} \BibitemShut
  {NoStop}%
\bibitem [{\citenamefont {Zeoli}\ \emph {et~al.}(2013)\citenamefont {Zeoli},
  \citenamefont {Machleidt},\ and\ \citenamefont {Entem}}]{ZE12}%
  \BibitemOpen
  \bibfield  {author} {\bibinfo {author} {\bibfnamefont {C.}~\bibnamefont
  {Zeoli}}, \bibinfo {author} {\bibfnamefont {R.}~\bibnamefont {Machleidt}}, \
  and\ \bibinfo {author} {\bibfnamefont {D.~R.}\ \bibnamefont {Entem}},\ }\href
  {\doibase 10.1007/s00601-012-0481-4} {\bibfield  {journal} {\bibinfo
  {journal} {Few Body Syst.}\ }\textbf {\bibinfo {volume} {54}},\ \bibinfo
  {pages} {2191} (\bibinfo {year} {2013})},\ \Eprint
  {http://arxiv.org/abs/1208.2657} {arXiv:1208.2657 [nucl-th]} \BibitemShut
  {NoStop}%
\bibitem [{\citenamefont {Birse}(2006)}]{Birse}%
  \BibitemOpen
  \bibfield  {author} {\bibinfo {author} {\bibfnamefont {M.~C.}\ \bibnamefont
  {Birse}},\ }\href {\doibase 10.1103/PhysRevC.74.014003} {\bibfield  {journal}
  {\bibinfo  {journal} {Phys. Rev. C}\ }\textbf {\bibinfo {volume} {74}},\
  \bibinfo {pages} {014003} (\bibinfo {year} {2006})},\ \Eprint
  {http://arxiv.org/abs/nucl-th/0507077} {arXiv:nucl-th/0507077} \BibitemShut
  {NoStop}%
\bibitem [{\citenamefont {Birse}(2007)}]{Birse:2007sx}%
  \BibitemOpen
  \bibfield  {author} {\bibinfo {author} {\bibfnamefont {M.~C.}\ \bibnamefont
  {Birse}},\ }\href {\doibase 10.1103/PhysRevC.76.034002} {\bibfield  {journal}
  {\bibinfo  {journal} {Phys. Rev. C}\ }\textbf {\bibinfo {volume} {76}},\
  \bibinfo {pages} {034002} (\bibinfo {year} {2007})},\ \Eprint
  {http://arxiv.org/abs/0706.0984} {arXiv:0706.0984 [nucl-th]} \BibitemShut
  {NoStop}%
\bibitem [{\citenamefont {Long}\ and\ \citenamefont {van
  Kolck}(2008)}]{Long:2007vp}%
  \BibitemOpen
  \bibfield  {author} {\bibinfo {author} {\bibfnamefont {B.}~\bibnamefont
  {Long}}\ and\ \bibinfo {author} {\bibfnamefont {U.}~\bibnamefont {van
  Kolck}},\ }\href {\doibase 10.1016/j.aop.2008.01.003} {\bibfield  {journal}
  {\bibinfo  {journal} {Annals Phys.}\ }\textbf {\bibinfo {volume} {323}},\
  \bibinfo {pages} {1304} (\bibinfo {year} {2008})},\ \Eprint
  {http://arxiv.org/abs/0707.4325} {arXiv:0707.4325 [quant-ph]} \BibitemShut
  {NoStop}%
\bibitem [{\citenamefont {Pav\'on~Valderrama}(2011{\natexlab{a}})}]{Valdper}%
  \BibitemOpen
  \bibfield  {author} {\bibinfo {author} {\bibfnamefont {M.}~\bibnamefont
  {Pav\'on~Valderrama}},\ }\href {\doibase 10.1103/PhysRevC.83.024003}
  {\bibfield  {journal} {\bibinfo  {journal} {Phys. Rev. C}\ }\textbf {\bibinfo
  {volume} {83}},\ \bibinfo {pages} {024003} (\bibinfo {year}
  {2011}{\natexlab{a}})},\ \Eprint {http://arxiv.org/abs/0912.0699}
  {arXiv:0912.0699 [nucl-th]} \BibitemShut {NoStop}%
\bibitem [{\citenamefont {Pav\'on~Valderrama}(2011{\natexlab{b}})}]{Valdperb}%
  \BibitemOpen
  \bibfield  {author} {\bibinfo {author} {\bibfnamefont {M.}~\bibnamefont
  {Pav\'on~Valderrama}},\ }\href {\doibase 10.1103/PhysRevC.84.064002}
  {\bibfield  {journal} {\bibinfo  {journal} {Phys. Rev. C}\ }\textbf {\bibinfo
  {volume} {84}},\ \bibinfo {pages} {064002} (\bibinfo {year}
  {2011}{\natexlab{b}})},\ \Eprint {http://arxiv.org/abs/1108.0872}
  {arXiv:1108.0872 [nucl-th]} \BibitemShut {NoStop}%
\bibitem [{\citenamefont {Long}\ and\ \citenamefont {Yang}(2011)}]{BY}%
  \BibitemOpen
  \bibfield  {author} {\bibinfo {author} {\bibfnamefont {B.}~\bibnamefont
  {Long}}\ and\ \bibinfo {author} {\bibfnamefont {C.~J.}\ \bibnamefont
  {Yang}},\ }\href {\doibase 10.1103/PhysRevC.84.057001} {\bibfield  {journal}
  {\bibinfo  {journal} {Phys. Rev. C}\ }\textbf {\bibinfo {volume} {84}},\
  \bibinfo {pages} {057001} (\bibinfo {year} {2011})},\ \Eprint
  {http://arxiv.org/abs/1108.0985} {arXiv:1108.0985 [nucl-th]} \BibitemShut
  {NoStop}%
\bibitem [{\citenamefont {Long}\ and\ \citenamefont
  {Yang}(2012{\natexlab{a}})}]{BYb}%
  \BibitemOpen
  \bibfield  {author} {\bibinfo {author} {\bibfnamefont {B.}~\bibnamefont
  {Long}}\ and\ \bibinfo {author} {\bibfnamefont {C.~J.}\ \bibnamefont
  {Yang}},\ }\href {\doibase 10.1103/PhysRevC.85.034002} {\bibfield  {journal}
  {\bibinfo  {journal} {Phys. Rev. C}\ }\textbf {\bibinfo {volume} {85}},\
  \bibinfo {pages} {034002} (\bibinfo {year} {2012}{\natexlab{a}})},\ \Eprint
  {http://arxiv.org/abs/1111.3993} {arXiv:1111.3993 [nucl-th]} \BibitemShut
  {NoStop}%
\bibitem [{\citenamefont {Long}\ and\ \citenamefont
  {Yang}(2012{\natexlab{b}})}]{BYc}%
  \BibitemOpen
  \bibfield  {author} {\bibinfo {author} {\bibfnamefont {B.}~\bibnamefont
  {Long}}\ and\ \bibinfo {author} {\bibfnamefont {C.~J.}\ \bibnamefont
  {Yang}},\ }\href {\doibase 10.1103/PhysRevC.86.024001} {\bibfield  {journal}
  {\bibinfo  {journal} {Phys. Rev. C}\ }\textbf {\bibinfo {volume} {86}},\
  \bibinfo {pages} {024001} (\bibinfo {year} {2012}{\natexlab{b}})},\ \Eprint
  {http://arxiv.org/abs/1202.4053} {arXiv:1202.4053 [nucl-th]} \BibitemShut
  {NoStop}%
\bibitem [{\citenamefont {Wu}\ and\ \citenamefont {Long}(2019)}]{bingwei18}%
  \BibitemOpen
  \bibfield  {author} {\bibinfo {author} {\bibfnamefont {S.}~\bibnamefont
  {Wu}}\ and\ \bibinfo {author} {\bibfnamefont {B.}~\bibnamefont {Long}},\
  }\href {\doibase 10.1103/PhysRevC.99.024003} {\bibfield  {journal} {\bibinfo
  {journal} {Phys. Rev. C}\ }\textbf {\bibinfo {volume} {99}},\ \bibinfo
  {pages} {024003} (\bibinfo {year} {2019})},\ \Eprint
  {http://arxiv.org/abs/1807.04407} {arXiv:1807.04407 [nucl-th]} \BibitemShut
  {NoStop}%
\bibitem [{\citenamefont {Song}\ \emph {et~al.}(2017)\citenamefont {Song},
  \citenamefont {Lazauskas},\ and\ \citenamefont {van Kolck}}]{Song:2016ale}%
  \BibitemOpen
  \bibfield  {author} {\bibinfo {author} {\bibfnamefont {Y.-H.}\ \bibnamefont
  {Song}}, \bibinfo {author} {\bibfnamefont {R.}~\bibnamefont {Lazauskas}}, \
  and\ \bibinfo {author} {\bibfnamefont {U.}~\bibnamefont {van Kolck}},\ }\href
  {\doibase 10.1103/PhysRevC.96.024002} {\bibfield  {journal} {\bibinfo
  {journal} {Phys. Rev. C}\ }\textbf {\bibinfo {volume} {96}},\ \bibinfo
  {pages} {024002} (\bibinfo {year} {2017})},\ \bibinfo {note} {[Erratum:
  Phys.Rev.C 100, 019901 (2019)]},\ \Eprint {http://arxiv.org/abs/1612.09090}
  {arXiv:1612.09090 [nucl-th]} \BibitemShut {NoStop}%
\bibitem [{\citenamefont {Yang}\ \emph {et~al.}(2021)\citenamefont {Yang},
  \citenamefont {Ekstr\"om}, \citenamefont {Forss\'en},\ and\ \citenamefont
  {Hagen}}]{Yang:2020pgi}%
  \BibitemOpen
  \bibfield  {author} {\bibinfo {author} {\bibfnamefont {C.-J.}\ \bibnamefont
  {Yang}}, \bibinfo {author} {\bibfnamefont {A.}~\bibnamefont {Ekstr\"om}},
  \bibinfo {author} {\bibfnamefont {C.}~\bibnamefont {Forss\'en}}, \ and\
  \bibinfo {author} {\bibfnamefont {G.}~\bibnamefont {Hagen}},\ }\href
  {\doibase 10.1103/PhysRevC.103.054304} {\bibfield  {journal} {\bibinfo
  {journal} {Phys. Rev. C}\ }\textbf {\bibinfo {volume} {103}},\ \bibinfo
  {pages} {054304} (\bibinfo {year} {2021})},\ \Eprint
  {http://arxiv.org/abs/2011.11584} {arXiv:2011.11584 [nucl-th]} \BibitemShut
  {NoStop}%
\bibitem [{\citenamefont {Stetcu}\ \emph {et~al.}(2007)\citenamefont {Stetcu},
  \citenamefont {Barrett},\ and\ \citenamefont {van Kolck}}]{Stetcu:2006ey}%
  \BibitemOpen
  \bibfield  {author} {\bibinfo {author} {\bibfnamefont {I.}~\bibnamefont
  {Stetcu}}, \bibinfo {author} {\bibfnamefont {B.~R.}\ \bibnamefont {Barrett}},
  \ and\ \bibinfo {author} {\bibfnamefont {U.}~\bibnamefont {van Kolck}},\
  }\href {\doibase 10.1016/j.physletb.2007.07.065} {\bibfield  {journal}
  {\bibinfo  {journal} {Phys. Lett. B}\ }\textbf {\bibinfo {volume} {653}},\
  \bibinfo {pages} {358} (\bibinfo {year} {2007})},\ \Eprint
  {http://arxiv.org/abs/nucl-th/0609023} {arXiv:nucl-th/0609023} \BibitemShut
  {NoStop}%
\bibitem [{\citenamefont {Contessi}\ \emph {et~al.}(2017)\citenamefont
  {Contessi}, \citenamefont {Lovato}, \citenamefont {Pederiva}, \citenamefont
  {Roggero}, \citenamefont {Kirscher},\ and\ \citenamefont {van
  Kolck}}]{Contessi:2017rww}%
  \BibitemOpen
  \bibfield  {author} {\bibinfo {author} {\bibfnamefont {L.}~\bibnamefont
  {Contessi}}, \bibinfo {author} {\bibfnamefont {A.}~\bibnamefont {Lovato}},
  \bibinfo {author} {\bibfnamefont {F.}~\bibnamefont {Pederiva}}, \bibinfo
  {author} {\bibfnamefont {A.}~\bibnamefont {Roggero}}, \bibinfo {author}
  {\bibfnamefont {J.}~\bibnamefont {Kirscher}}, \ and\ \bibinfo {author}
  {\bibfnamefont {U.}~\bibnamefont {van Kolck}},\ }\href {\doibase
  10.1016/j.physletb.2017.07.048} {\bibfield  {journal} {\bibinfo  {journal}
  {Phys. Lett. B}\ }\textbf {\bibinfo {volume} {772}},\ \bibinfo {pages} {839}
  (\bibinfo {year} {2017})},\ \Eprint {http://arxiv.org/abs/1701.06516}
  {arXiv:1701.06516 [nucl-th]} \BibitemShut {NoStop}%
\bibitem [{\citenamefont {Bansal}\ \emph {et~al.}(2018)\citenamefont {Bansal},
  \citenamefont {Binder}, \citenamefont {Ekstr\"om}, \citenamefont {Hagen},
  \citenamefont {Jansen},\ and\ \citenamefont {Papenbrock}}]{pionless16b}%
  \BibitemOpen
  \bibfield  {author} {\bibinfo {author} {\bibfnamefont {A.}~\bibnamefont
  {Bansal}}, \bibinfo {author} {\bibfnamefont {S.}~\bibnamefont {Binder}},
  \bibinfo {author} {\bibfnamefont {A.}~\bibnamefont {Ekstr\"om}}, \bibinfo
  {author} {\bibfnamefont {G.}~\bibnamefont {Hagen}}, \bibinfo {author}
  {\bibfnamefont {G.~R.}\ \bibnamefont {Jansen}}, \ and\ \bibinfo {author}
  {\bibfnamefont {T.}~\bibnamefont {Papenbrock}},\ }\href {\doibase
  10.1103/PhysRevC.98.054301} {\bibfield  {journal} {\bibinfo  {journal} {Phys.
  Rev. C}\ }\textbf {\bibinfo {volume} {98}},\ \bibinfo {pages} {054301}
  (\bibinfo {year} {2018})},\ \Eprint {http://arxiv.org/abs/1712.10246}
  {arXiv:1712.10246 [nucl-th]} \BibitemShut {NoStop}%
\bibitem [{\citenamefont {Birse}(2010)}]{Birse:2010jr}%
  \BibitemOpen
  \bibfield  {author} {\bibinfo {author} {\bibfnamefont {M.~C.}\ \bibnamefont
  {Birse}},\ }\href {\doibase 10.1140/epja/i2010-11034-9} {\bibfield  {journal}
  {\bibinfo  {journal} {Eur. Phys. J. A}\ }\textbf {\bibinfo {volume} {46}},\
  \bibinfo {pages} {231} (\bibinfo {year} {2010})},\ \Eprint
  {http://arxiv.org/abs/1007.0540} {arXiv:1007.0540 [nucl-th]} \BibitemShut
  {NoStop}%
\bibitem [{\citenamefont {Long}(2013)}]{Bs}%
  \BibitemOpen
  \bibfield  {author} {\bibinfo {author} {\bibfnamefont {B.}~\bibnamefont
  {Long}},\ }\href {\doibase 10.1103/PhysRevC.88.014002} {\bibfield  {journal}
  {\bibinfo  {journal} {Phys. Rev. C}\ }\textbf {\bibinfo {volume} {88}},\
  \bibinfo {pages} {014002} (\bibinfo {year} {2013})},\ \Eprint
  {http://arxiv.org/abs/1304.7382} {arXiv:1304.7382 [nucl-th]} \BibitemShut
  {NoStop}%
\bibitem [{\citenamefont {S\'anchez~S\'anchez}\ \emph
  {et~al.}(2018)\citenamefont {S\'anchez~S\'anchez}, \citenamefont {Yang},
  \citenamefont {Long},\ and\ \citenamefont {van
  Kolck}}]{SanchezSanchez:2017tws}%
  \BibitemOpen
  \bibfield  {author} {\bibinfo {author} {\bibfnamefont {M.}~\bibnamefont
  {S\'anchez~S\'anchez}}, \bibinfo {author} {\bibfnamefont {C.-J.}\
  \bibnamefont {Yang}}, \bibinfo {author} {\bibfnamefont {B.}~\bibnamefont
  {Long}}, \ and\ \bibinfo {author} {\bibfnamefont {U.}~\bibnamefont {van
  Kolck}},\ }\href {\doibase 10.1103/PhysRevC.97.024001} {\bibfield  {journal}
  {\bibinfo  {journal} {Phys. Rev. C}\ }\textbf {\bibinfo {volume} {97}},\
  \bibinfo {pages} {024001} (\bibinfo {year} {2018})},\ \Eprint
  {http://arxiv.org/abs/1704.08524} {arXiv:1704.08524 [nucl-th]} \BibitemShut
  {NoStop}%
\bibitem [{\citenamefont {Bedaque}\ \emph {et~al.}(2000)\citenamefont
  {Bedaque}, \citenamefont {Hammer},\ and\ \citenamefont {van
  Kolck}}]{Bedaque:1999ve}%
  \BibitemOpen
  \bibfield  {author} {\bibinfo {author} {\bibfnamefont {P.~F.}\ \bibnamefont
  {Bedaque}}, \bibinfo {author} {\bibfnamefont {H.~W.}\ \bibnamefont {Hammer}},
  \ and\ \bibinfo {author} {\bibfnamefont {U.}~\bibnamefont {van Kolck}},\
  }\href {\doibase 10.1016/S0375-9474(00)00205-0} {\bibfield  {journal}
  {\bibinfo  {journal} {Nucl. Phys. A}\ }\textbf {\bibinfo {volume} {676}},\
  \bibinfo {pages} {357} (\bibinfo {year} {2000})},\ \Eprint
  {http://arxiv.org/abs/nucl-th/9906032} {arXiv:nucl-th/9906032} \BibitemShut
  {NoStop}%
\bibitem [{\citenamefont {Kievsky}\ \emph {et~al.}(2017)\citenamefont
  {Kievsky}, \citenamefont {Viviani}, \citenamefont {Gattobigio},\ and\
  \citenamefont {Girlanda}}]{Pisa}%
  \BibitemOpen
  \bibfield  {author} {\bibinfo {author} {\bibfnamefont {A.}~\bibnamefont
  {Kievsky}}, \bibinfo {author} {\bibfnamefont {M.}~\bibnamefont {Viviani}},
  \bibinfo {author} {\bibfnamefont {M.}~\bibnamefont {Gattobigio}}, \ and\
  \bibinfo {author} {\bibfnamefont {L.}~\bibnamefont {Girlanda}},\ }\href
  {\doibase 10.1103/PhysRevC.95.024001} {\bibfield  {journal} {\bibinfo
  {journal} {Phys. Rev. C}\ }\textbf {\bibinfo {volume} {95}},\ \bibinfo
  {pages} {024001} (\bibinfo {year} {2017})},\ \Eprint
  {http://arxiv.org/abs/1610.09858} {arXiv:1610.09858 [nucl-th]} \BibitemShut
  {NoStop}%
\bibitem [{\citenamefont {Kievsky}\ \emph {et~al.}(2018)\citenamefont
  {Kievsky}, \citenamefont {Viviani}, \citenamefont {Logoteta}, \citenamefont
  {Bombaci},\ and\ \citenamefont {Girlanda}}]{Kievsky:2018xsl}%
  \BibitemOpen
  \bibfield  {author} {\bibinfo {author} {\bibfnamefont {A.}~\bibnamefont
  {Kievsky}}, \bibinfo {author} {\bibfnamefont {M.}~\bibnamefont {Viviani}},
  \bibinfo {author} {\bibfnamefont {D.}~\bibnamefont {Logoteta}}, \bibinfo
  {author} {\bibfnamefont {I.}~\bibnamefont {Bombaci}}, \ and\ \bibinfo
  {author} {\bibfnamefont {L.}~\bibnamefont {Girlanda}},\ }\href {\doibase
  10.1103/PhysRevLett.121.072701} {\bibfield  {journal} {\bibinfo  {journal}
  {Phys. Rev. Lett.}\ }\textbf {\bibinfo {volume} {121}},\ \bibinfo {pages}
  {072701} (\bibinfo {year} {2018})},\ \Eprint
  {http://arxiv.org/abs/1806.02636} {arXiv:1806.02636 [nucl-th]} \BibitemShut
  {NoStop}%
\bibitem [{\citenamefont {van Kolck}(2020{\natexlab{b}})}]{vanKolck:2020plz}%
  \BibitemOpen
  \bibfield  {author} {\bibinfo {author} {\bibfnamefont {U.}~\bibnamefont {van
  Kolck}},\ }\href {\doibase 10.1140/epja/s10050-020-00092-1} {\bibfield
  {journal} {\bibinfo  {journal} {Eur. Phys. J. A}\ }\textbf {\bibinfo {volume}
  {56}},\ \bibinfo {pages} {97} (\bibinfo {year} {2020}{\natexlab{b}})},\
  \Eprint {http://arxiv.org/abs/2003.09974} {arXiv:2003.09974 [nucl-th]}
  \BibitemShut {NoStop}%
\bibitem [{\citenamefont {Yang}(2020)}]{yang_rev}%
  \BibitemOpen
  \bibfield  {author} {\bibinfo {author} {\bibfnamefont {C.~J.}\ \bibnamefont
  {Yang}},\ }\href {\doibase 10.1140/epja/s10050-020-00104-0} {\bibfield
  {journal} {\bibinfo  {journal} {Eur. Phys. J. A}\ }\textbf {\bibinfo {volume}
  {56}},\ \bibinfo {pages} {96} (\bibinfo {year} {2020})},\ \Eprint
  {http://arxiv.org/abs/1905.12510} {arXiv:1905.12510 [nucl-th]} \BibitemShut
  {NoStop}%
\bibitem [{\citenamefont {Manohar}\ and\ \citenamefont
  {Georgi}(1984)}]{Manohar:1984}%
  \BibitemOpen
  \bibfield  {author} {\bibinfo {author} {\bibfnamefont {A.}~\bibnamefont
  {Manohar}}\ and\ \bibinfo {author} {\bibfnamefont {H.}~\bibnamefont
  {Georgi}},\ }\href {\doibase 10.1016/0550-3213(84)90231-1} {\bibfield
  {journal} {\bibinfo  {journal} {Nucl. Phys. B}\ }\textbf {\bibinfo {volume}
  {234}},\ \bibinfo {pages} {189} (\bibinfo {year} {1984})}\BibitemShut
  {NoStop}%
\bibitem [{\citenamefont {Georgi}\ and\ \citenamefont {Randall}(1986)}]{nda2}%
  \BibitemOpen
  \bibfield  {author} {\bibinfo {author} {\bibfnamefont {H.}~\bibnamefont
  {Georgi}}\ and\ \bibinfo {author} {\bibfnamefont {L.}~\bibnamefont
  {Randall}},\ }\href {\doibase 10.1016/0550-3213(86)90022-2} {\bibfield
  {journal} {\bibinfo  {journal} {Nucl. Phys. B}\ }\textbf {\bibinfo {volume}
  {276}},\ \bibinfo {pages} {241} (\bibinfo {year} {1986})}\BibitemShut
  {NoStop}%
\bibitem [{\citenamefont {Weinberg}(1989)}]{nda3}%
  \BibitemOpen
  \bibfield  {author} {\bibinfo {author} {\bibfnamefont {S.}~\bibnamefont
  {Weinberg}},\ }\href {\doibase 10.1103/physrevlett.63.2333} {\bibfield
  {journal} {\bibinfo  {journal} {Phys. Rev. Lett.}\ }\textbf {\bibinfo
  {volume} {63}},\ \bibinfo {pages} {2333} (\bibinfo {year}
  {1989})}\BibitemShut {NoStop}%
\bibitem [{\citenamefont {Georgi}(1993)}]{nda4}%
  \BibitemOpen
  \bibfield  {author} {\bibinfo {author} {\bibfnamefont {H.}~\bibnamefont
  {Georgi}},\ }\href {\doibase 10.1016/0370-2693(93)91728-6} {\bibfield
  {journal} {\bibinfo  {journal} {Phys. Lett. B}\ }\textbf {\bibinfo {volume}
  {298}},\ \bibinfo {pages} {187} (\bibinfo {year} {1993})}\BibitemShut
  {NoStop}%
\bibitem [{\citenamefont {van Kolck}(1994)}]{vanKolck:1994yi}%
  \BibitemOpen
  \bibfield  {author} {\bibinfo {author} {\bibfnamefont {U.}~\bibnamefont {van
  Kolck}},\ }\href {\doibase 10.1103/PhysRevC.49.2932} {\bibfield  {journal}
  {\bibinfo  {journal} {Phys. Rev. C}\ }\textbf {\bibinfo {volume} {49}},\
  \bibinfo {pages} {2932} (\bibinfo {year} {1994})}\BibitemShut {NoStop}%
\bibitem [{\citenamefont {Epelbaum}\ \emph {et~al.}(2002)\citenamefont
  {Epelbaum}, \citenamefont {Nogga}, \citenamefont {Gl\"ockle}, \citenamefont
  {Kamada}, \citenamefont {Mei\ss{}ner},\ and\ \citenamefont
  {Wita\l{}a}}]{PhysRevC.66.064001}%
  \BibitemOpen
  \bibfield  {author} {\bibinfo {author} {\bibfnamefont {E.}~\bibnamefont
  {Epelbaum}}, \bibinfo {author} {\bibfnamefont {A.}~\bibnamefont {Nogga}},
  \bibinfo {author} {\bibfnamefont {W.}~\bibnamefont {Gl\"ockle}}, \bibinfo
  {author} {\bibfnamefont {H.}~\bibnamefont {Kamada}}, \bibinfo {author}
  {\bibfnamefont {U.-G.}\ \bibnamefont {Mei\ss{}ner}}, \ and\ \bibinfo {author}
  {\bibfnamefont {H.}~\bibnamefont {Wita\l{}a}},\ }\href {\doibase
  10.1103/PhysRevC.66.064001} {\bibfield  {journal} {\bibinfo  {journal} {Phys.
  Rev. C}\ }\textbf {\bibinfo {volume} {66}},\ \bibinfo {pages} {064001}
  (\bibinfo {year} {2002})}\BibitemShut {NoStop}%
\bibitem [{\citenamefont {Navratil}\ \emph {et~al.}(2000)\citenamefont
  {Navratil}, \citenamefont {Kamuntavicius},\ and\ \citenamefont
  {Barrett}}]{Navratil:1999pw}%
  \BibitemOpen
  \bibfield  {author} {\bibinfo {author} {\bibfnamefont {P.}~\bibnamefont
  {Navratil}}, \bibinfo {author} {\bibfnamefont {G.~P.}\ \bibnamefont
  {Kamuntavicius}}, \ and\ \bibinfo {author} {\bibfnamefont {B.~R.}\
  \bibnamefont {Barrett}},\ }\href {\doibase 10.1103/PhysRevC.61.044001}
  {\bibfield  {journal} {\bibinfo  {journal} {Phys. Rev. C}\ }\textbf {\bibinfo
  {volume} {61}},\ \bibinfo {pages} {044001} (\bibinfo {year} {2000})},\
  \Eprint {http://arxiv.org/abs/nucl-th/9907054} {arXiv:nucl-th/9907054}
  \BibitemShut {NoStop}%
\bibitem [{\citenamefont {Friar}(1997)}]{Friar1997}%
  \BibitemOpen
  \bibfield  {author} {\bibinfo {author} {\bibfnamefont {J.~L.}\ \bibnamefont
  {Friar}},\ }\href {\doibase 10.1007/s006010050059} {\bibfield  {journal}
  {\bibinfo  {journal} {Few-Body Systems}\ }\textbf {\bibinfo {volume} {22}},\
  \bibinfo {pages} {161} (\bibinfo {year} {1997})}\BibitemShut {NoStop}%
\bibitem [{\citenamefont {Epelbaum}(2006)}]{Epelbaum:2006eu}%
  \BibitemOpen
  \bibfield  {author} {\bibinfo {author} {\bibfnamefont {E.}~\bibnamefont
  {Epelbaum}},\ }\href {\doibase 10.1016/j.physletb.2006.06.046} {\bibfield
  {journal} {\bibinfo  {journal} {Phys. Lett. B}\ }\textbf {\bibinfo {volume}
  {639}},\ \bibinfo {pages} {456} (\bibinfo {year} {2006})},\ \Eprint
  {http://arxiv.org/abs/nucl-th/0511025} {arXiv:nucl-th/0511025} \BibitemShut
  {NoStop}%
\bibitem [{\citenamefont {Bergervoet}\ \emph {et~al.}(1990)\citenamefont
  {Bergervoet}, \citenamefont {van Campen}, \citenamefont {Klomp},
  \citenamefont {de~Kok}, \citenamefont {Rijken}, \citenamefont {Stoks},\ and\
  \citenamefont {de~Swart}}]{nij}%
  \BibitemOpen
  \bibfield  {author} {\bibinfo {author} {\bibfnamefont {J.~R.}\ \bibnamefont
  {Bergervoet}}, \bibinfo {author} {\bibfnamefont {P.~C.}\ \bibnamefont {van
  Campen}}, \bibinfo {author} {\bibfnamefont {R.~A.~M.}\ \bibnamefont {Klomp}},
  \bibinfo {author} {\bibfnamefont {J.-L.}\ \bibnamefont {de~Kok}}, \bibinfo
  {author} {\bibfnamefont {T.~A.}\ \bibnamefont {Rijken}}, \bibinfo {author}
  {\bibfnamefont {V.~G.~J.}\ \bibnamefont {Stoks}}, \ and\ \bibinfo {author}
  {\bibfnamefont {J.~J.}\ \bibnamefont {de~Swart}},\ }\href {\doibase
  10.1103/physrevc.41.1435} {\bibfield  {journal} {\bibinfo  {journal} {Phys.
  Rev. C}\ }\textbf {\bibinfo {volume} {41}},\ \bibinfo {pages} {1435}
  (\bibinfo {year} {1990})}\BibitemShut {NoStop}%
\bibitem [{\citenamefont {Navr{\'{a}}til}\ \emph {et~al.}(2000)\citenamefont
  {Navr{\'{a}}til}, \citenamefont {Vary},\ and\ \citenamefont
  {Barrett}}]{ncsm}%
  \BibitemOpen
  \bibfield  {author} {\bibinfo {author} {\bibfnamefont {P.}~\bibnamefont
  {Navr{\'{a}}til}}, \bibinfo {author} {\bibfnamefont {J.~P.}\ \bibnamefont
  {Vary}}, \ and\ \bibinfo {author} {\bibfnamefont {B.~R.}\ \bibnamefont
  {Barrett}},\ }\href {\doibase 10.1103/physrevlett.84.5728} {\bibfield
  {journal} {\bibinfo  {journal} {Phys. Rev. Lett.}\ }\textbf {\bibinfo
  {volume} {84}},\ \bibinfo {pages} {5728} (\bibinfo {year}
  {2000})}\BibitemShut {NoStop}%
\bibitem [{\citenamefont {Navratil}\ \emph {et~al.}(2000)\citenamefont
  {Navratil}, \citenamefont {Vary},\ and\ \citenamefont {Barrett}}]{ncsma}%
  \BibitemOpen
  \bibfield  {author} {\bibinfo {author} {\bibfnamefont {P.}~\bibnamefont
  {Navratil}}, \bibinfo {author} {\bibfnamefont {J.~P.}\ \bibnamefont {Vary}},
  \ and\ \bibinfo {author} {\bibfnamefont {B.~R.}\ \bibnamefont {Barrett}},\
  }\href {\doibase 10.1103/PhysRevC.62.054311} {\bibfield  {journal} {\bibinfo
  {journal} {Phys. Rev. C}\ }\textbf {\bibinfo {volume} {62}},\ \bibinfo
  {pages} {054311} (\bibinfo {year} {2000})}\BibitemShut {NoStop}%
\bibitem [{\citenamefont {K{\"u}mmel}\ \emph {et~al.}(1978)\citenamefont
  {K{\"u}mmel}, \citenamefont {L{\"u}hrmann},\ and\ \citenamefont
  {Zabolitzky}}]{kuemmel1978}%
  \BibitemOpen
  \bibfield  {author} {\bibinfo {author} {\bibfnamefont {H.}~\bibnamefont
  {K{\"u}mmel}}, \bibinfo {author} {\bibfnamefont {K.~H.}\ \bibnamefont
  {L{\"u}hrmann}}, \ and\ \bibinfo {author} {\bibfnamefont {J.~G.}\
  \bibnamefont {Zabolitzky}},\ }\href {\doibase 10.1016/0370-1573(78)90081-9}
  {\bibfield  {journal} {\bibinfo  {journal} {Physics Reports}\ }\textbf
  {\bibinfo {volume} {36}},\ \bibinfo {pages} {1 } (\bibinfo {year}
  {1978})}\BibitemShut {NoStop}%
\bibitem [{\citenamefont {Bartlett}\ and\ \citenamefont
  {Musia{\l}}(2007)}]{cc}%
  \BibitemOpen
  \bibfield  {author} {\bibinfo {author} {\bibfnamefont {R.~J.}\ \bibnamefont
  {Bartlett}}\ and\ \bibinfo {author} {\bibfnamefont {M.}~\bibnamefont
  {Musia{\l}}},\ }\href {\doibase 10.1103/revmodphys.79.291} {\bibfield
  {journal} {\bibinfo  {journal} {Rev. Mod. Phys.}\ }\textbf {\bibinfo {volume}
  {79}},\ \bibinfo {pages} {291} (\bibinfo {year} {2007})}\BibitemShut
  {NoStop}%
\bibitem [{\citenamefont {Hagen}\ \emph
  {et~al.}(2014{\natexlab{a}})\citenamefont {Hagen}, \citenamefont
  {Papenbrock}, \citenamefont {Hjorth-Jensen},\ and\ \citenamefont
  {Dean}}]{hagen2014}%
  \BibitemOpen
  \bibfield  {author} {\bibinfo {author} {\bibfnamefont {G.}~\bibnamefont
  {Hagen}}, \bibinfo {author} {\bibfnamefont {T.}~\bibnamefont {Papenbrock}},
  \bibinfo {author} {\bibfnamefont {M.}~\bibnamefont {Hjorth-Jensen}}, \ and\
  \bibinfo {author} {\bibfnamefont {D.~J.}\ \bibnamefont {Dean}},\ }\href
  {\doibase 10.1088/0034-4885/77/9/096302} {\bibfield  {journal} {\bibinfo
  {journal} {Rep. Prog. Phys.}\ }\textbf {\bibinfo {volume} {77}},\ \bibinfo
  {pages} {096302} (\bibinfo {year} {2014}{\natexlab{a}})}\BibitemShut
  {NoStop}%
\bibitem [{\citenamefont {Hagen}\ \emph {et~al.}(2007)\citenamefont {Hagen},
  \citenamefont {Papenbrock}, \citenamefont {Dean}, \citenamefont {Schwenk},
  \citenamefont {Nogga}, \citenamefont {Wloch},\ and\ \citenamefont
  {Piecuch}}]{normal1}%
  \BibitemOpen
  \bibfield  {author} {\bibinfo {author} {\bibfnamefont {G.}~\bibnamefont
  {Hagen}}, \bibinfo {author} {\bibfnamefont {T.}~\bibnamefont {Papenbrock}},
  \bibinfo {author} {\bibfnamefont {D.}~\bibnamefont {Dean}}, \bibinfo {author}
  {\bibfnamefont {A.}~\bibnamefont {Schwenk}}, \bibinfo {author} {\bibfnamefont
  {A.}~\bibnamefont {Nogga}}, \bibinfo {author} {\bibfnamefont
  {M.}~\bibnamefont {Wloch}}, \ and\ \bibinfo {author} {\bibfnamefont
  {P.}~\bibnamefont {Piecuch}},\ }\href {\doibase 10.1103/PhysRevC.76.034302}
  {\bibfield  {journal} {\bibinfo  {journal} {Phys. Rev. C}\ }\textbf {\bibinfo
  {volume} {76}},\ \bibinfo {pages} {034302} (\bibinfo {year} {2007})},\
  \Eprint {http://arxiv.org/abs/0704.2854} {arXiv:0704.2854 [nucl-th]}
  \BibitemShut {NoStop}%
\bibitem [{\citenamefont {Binder}\ \emph {et~al.}(2013)\citenamefont {Binder},
  \citenamefont {Langhammer}, \citenamefont {Calci}, \citenamefont {Navratil},\
  and\ \citenamefont {Roth}}]{normal2}%
  \BibitemOpen
  \bibfield  {author} {\bibinfo {author} {\bibfnamefont {S.}~\bibnamefont
  {Binder}}, \bibinfo {author} {\bibfnamefont {J.}~\bibnamefont {Langhammer}},
  \bibinfo {author} {\bibfnamefont {A.}~\bibnamefont {Calci}}, \bibinfo
  {author} {\bibfnamefont {P.}~\bibnamefont {Navratil}}, \ and\ \bibinfo
  {author} {\bibfnamefont {R.}~\bibnamefont {Roth}},\ }\href {\doibase
  10.1103/PhysRevC.87.021303} {\bibfield  {journal} {\bibinfo  {journal} {Phys.
  Rev. C}\ }\textbf {\bibinfo {volume} {87}},\ \bibinfo {pages} {021303}
  (\bibinfo {year} {2013})},\ \Eprint {http://arxiv.org/abs/1211.4748}
  {arXiv:1211.4748 [nucl-th]} \BibitemShut {NoStop}%
\bibitem [{\citenamefont {Taube}\ and\ \citenamefont
  {Bartlett}(2008)}]{taube2008}%
  \BibitemOpen
  \bibfield  {author} {\bibinfo {author} {\bibfnamefont {A.~G.}\ \bibnamefont
  {Taube}}\ and\ \bibinfo {author} {\bibfnamefont {R.~J.}\ \bibnamefont
  {Bartlett}},\ }\href {\doibase 10.1063/1.2830236} {\bibfield  {journal}
  {\bibinfo  {journal} {J. Chem. Phys.}\ }\textbf {\bibinfo {volume} {128}},\
  \bibinfo {eid} {044110} (\bibinfo {year} {2008})}\BibitemShut {NoStop}%
\bibitem [{\citenamefont {Hagen}\ \emph
  {et~al.}(2014{\natexlab{b}})\citenamefont {Hagen}, \citenamefont
  {Papenbrock}, \citenamefont {Ekstr\"{o}m}, \citenamefont {Wendt},
  \citenamefont {Baardsen}, \citenamefont {Gandolfi}, \citenamefont
  {Hjorth-Jensen},\ and\ \citenamefont {Horowitz}}]{cc_eos2}%
  \BibitemOpen
  \bibfield  {author} {\bibinfo {author} {\bibfnamefont {G.}~\bibnamefont
  {Hagen}}, \bibinfo {author} {\bibfnamefont {T.}~\bibnamefont {Papenbrock}},
  \bibinfo {author} {\bibfnamefont {A.}~\bibnamefont {Ekstr\"{o}m}}, \bibinfo
  {author} {\bibfnamefont {K.~A.}\ \bibnamefont {Wendt}}, \bibinfo {author}
  {\bibfnamefont {G.}~\bibnamefont {Baardsen}}, \bibinfo {author}
  {\bibfnamefont {S.}~\bibnamefont {Gandolfi}}, \bibinfo {author}
  {\bibfnamefont {M.}~\bibnamefont {Hjorth-Jensen}}, \ and\ \bibinfo {author}
  {\bibfnamefont {C.~J.}\ \bibnamefont {Horowitz}},\ }\href {\doibase
  10.1103/physrevc.89.014319} {\bibfield  {journal} {\bibinfo  {journal} {Phys.
  Rev. C}\ }\textbf {\bibinfo {volume} {89}} (\bibinfo {year}
  {2014}{\natexlab{b}}),\ 10.1103/physrevc.89.014319}\BibitemShut {NoStop}%
\bibitem [{\citenamefont {Hoferichter}\ \emph {et~al.}(2015)\citenamefont
  {Hoferichter}, \citenamefont {Ruiz~de Elvira}, \citenamefont {Kubis},\ and\
  \citenamefont {Mei\ss{}ner}}]{Hoferichter:2015tha}%
  \BibitemOpen
  \bibfield  {author} {\bibinfo {author} {\bibfnamefont {M.}~\bibnamefont
  {Hoferichter}}, \bibinfo {author} {\bibfnamefont {J.}~\bibnamefont {Ruiz~de
  Elvira}}, \bibinfo {author} {\bibfnamefont {B.}~\bibnamefont {Kubis}}, \ and\
  \bibinfo {author} {\bibfnamefont {U.-G.}\ \bibnamefont {Mei\ss{}ner}},\
  }\href {\doibase 10.1103/PhysRevLett.115.192301} {\bibfield  {journal}
  {\bibinfo  {journal} {Phys. Rev. Lett.}\ }\textbf {\bibinfo {volume} {115}},\
  \bibinfo {pages} {192301} (\bibinfo {year} {2015})},\ \Eprint
  {http://arxiv.org/abs/1507.07552} {arXiv:1507.07552 [nucl-th]} \BibitemShut
  {NoStop}%
\bibitem [{\citenamefont {Fujita}\ and\ \citenamefont
  {Miyazawa}(1957)}]{Fujita1957}%
  \BibitemOpen
  \bibfield  {author} {\bibinfo {author} {\bibfnamefont {J.}~\bibnamefont
  {Fujita}}\ and\ \bibinfo {author} {\bibfnamefont {H.}~\bibnamefont
  {Miyazawa}},\ }\href {\doibase 10.1143/PTP.17.360} {\bibfield  {journal}
  {\bibinfo  {journal} {Prog. Theor. Phys.}\ }\textbf {\bibinfo {volume}
  {17}},\ \bibinfo {pages} {360} (\bibinfo {year} {1957})}\BibitemShut
  {NoStop}%
\bibitem [{\citenamefont {Platter}\ \emph {et~al.}(2005)\citenamefont
  {Platter}, \citenamefont {Hammer},\ and\ \citenamefont
  {Meissner}}]{Platter:2004zs}%
  \BibitemOpen
  \bibfield  {author} {\bibinfo {author} {\bibfnamefont {L.}~\bibnamefont
  {Platter}}, \bibinfo {author} {\bibfnamefont {H.~W.}\ \bibnamefont {Hammer}},
  \ and\ \bibinfo {author} {\bibfnamefont {U.-G.}\ \bibnamefont {Meissner}},\
  }\href {\doibase 10.1016/j.physletb.2004.12.068} {\bibfield  {journal}
  {\bibinfo  {journal} {Phys. Lett. B}\ }\textbf {\bibinfo {volume} {607}},\
  \bibinfo {pages} {254} (\bibinfo {year} {2005})},\ \Eprint
  {http://arxiv.org/abs/nucl-th/0409040} {arXiv:nucl-th/0409040} \BibitemShut
  {NoStop}%
\bibitem [{\citenamefont {Navratil}\ \emph {et~al.}(2007)\citenamefont
  {Navratil}, \citenamefont {Gueorguiev}, \citenamefont {Vary}, \citenamefont
  {Ormand},\ and\ \citenamefont {Nogga}}]{corr}%
  \BibitemOpen
  \bibfield  {author} {\bibinfo {author} {\bibfnamefont {P.}~\bibnamefont
  {Navratil}}, \bibinfo {author} {\bibfnamefont {V.~G.}\ \bibnamefont
  {Gueorguiev}}, \bibinfo {author} {\bibfnamefont {J.~P.}\ \bibnamefont
  {Vary}}, \bibinfo {author} {\bibfnamefont {W.~E.}\ \bibnamefont {Ormand}}, \
  and\ \bibinfo {author} {\bibfnamefont {A.}~\bibnamefont {Nogga}},\ }\href
  {\doibase 10.1103/PhysRevLett.99.042501} {\bibfield  {journal} {\bibinfo
  {journal} {Phys. Rev. Lett.}\ }\textbf {\bibinfo {volume} {99}},\ \bibinfo
  {pages} {042501} (\bibinfo {year} {2007})},\ \Eprint
  {http://arxiv.org/abs/nucl-th/0701038} {arXiv:nucl-th/0701038} \BibitemShut
  {NoStop}%
\bibitem [{\citenamefont {Nogga}\ \emph {et~al.}(2006)\citenamefont {Nogga},
  \citenamefont {Navratil}, \citenamefont {Barrett},\ and\ \citenamefont
  {Vary}}]{corr1}%
  \BibitemOpen
  \bibfield  {author} {\bibinfo {author} {\bibfnamefont {A.}~\bibnamefont
  {Nogga}}, \bibinfo {author} {\bibfnamefont {P.}~\bibnamefont {Navratil}},
  \bibinfo {author} {\bibfnamefont {B.~R.}\ \bibnamefont {Barrett}}, \ and\
  \bibinfo {author} {\bibfnamefont {J.~P.}\ \bibnamefont {Vary}},\ }\href
  {\doibase 10.1103/PhysRevC.73.064002} {\bibfield  {journal} {\bibinfo
  {journal} {Phys. Rev. C}\ }\textbf {\bibinfo {volume} {73}},\ \bibinfo
  {pages} {064002} (\bibinfo {year} {2006})},\ \Eprint
  {http://arxiv.org/abs/nucl-th/0511082} {arXiv:nucl-th/0511082} \BibitemShut
  {NoStop}%
\bibitem [{\citenamefont {Wesolowski}\ \emph {et~al.}(2021)\citenamefont
  {Wesolowski}, \citenamefont {Svensson}, \citenamefont {Ekström},
  \citenamefont {Forssén}, \citenamefont {Furnstahl}, \citenamefont
  {Melendez},\ and\ \citenamefont {Phillips}}]{wesolowski2021fast}%
  \BibitemOpen
  \bibfield  {author} {\bibinfo {author} {\bibfnamefont {S.}~\bibnamefont
  {Wesolowski}}, \bibinfo {author} {\bibfnamefont {I.}~\bibnamefont
  {Svensson}}, \bibinfo {author} {\bibfnamefont {A.}~\bibnamefont {Ekström}},
  \bibinfo {author} {\bibfnamefont {C.}~\bibnamefont {Forssén}}, \bibinfo
  {author} {\bibfnamefont {R.~J.}\ \bibnamefont {Furnstahl}}, \bibinfo {author}
  {\bibfnamefont {J.~A.}\ \bibnamefont {Melendez}}, \ and\ \bibinfo {author}
  {\bibfnamefont {D.~R.}\ \bibnamefont {Phillips}},\ }\href@noop {} {\enquote
  {\bibinfo {title} {{Fast \& rigorous constraints on chiral three-nucleon
  forces from few-body observables}},}\ } (\bibinfo {year} {2021}),\ \Eprint
  {http://arxiv.org/abs/2104.04441} {arXiv:2104.04441 [nucl-th]} \BibitemShut
  {NoStop}%
\bibitem [{\citenamefont {L\"ahde}\ \emph {et~al.}(2014)\citenamefont
  {L\"ahde}, \citenamefont {Epelbaum}, \citenamefont {Krebs}, \citenamefont
  {Lee}, \citenamefont {Mei\ss{}ner},\ and\ \citenamefont
  {Rupak}}]{Lahde:2013uqa}%
  \BibitemOpen
  \bibfield  {author} {\bibinfo {author} {\bibfnamefont {T.~A.}\ \bibnamefont
  {L\"ahde}}, \bibinfo {author} {\bibfnamefont {E.}~\bibnamefont {Epelbaum}},
  \bibinfo {author} {\bibfnamefont {H.}~\bibnamefont {Krebs}}, \bibinfo
  {author} {\bibfnamefont {D.}~\bibnamefont {Lee}}, \bibinfo {author}
  {\bibfnamefont {U.-G.}\ \bibnamefont {Mei\ss{}ner}}, \ and\ \bibinfo {author}
  {\bibfnamefont {G.}~\bibnamefont {Rupak}},\ }\href {\doibase
  10.1016/j.physletb.2014.03.023} {\bibfield  {journal} {\bibinfo  {journal}
  {Phys. Lett. B}\ }\textbf {\bibinfo {volume} {732}},\ \bibinfo {pages} {110}
  (\bibinfo {year} {2014})},\ \Eprint {http://arxiv.org/abs/1311.0477}
  {arXiv:1311.0477 [nucl-th]} \BibitemShut {NoStop}%
\bibitem [{\citenamefont {Machleidt}\ \emph {et~al.}(2010)\citenamefont
  {Machleidt}, \citenamefont {Liu}, \citenamefont {Entem},\ and\ \citenamefont
  {Ruiz~Arriola}}]{Machleidt:2009bh}%
  \BibitemOpen
  \bibfield  {author} {\bibinfo {author} {\bibfnamefont {R.}~\bibnamefont
  {Machleidt}}, \bibinfo {author} {\bibfnamefont {P.}~\bibnamefont {Liu}},
  \bibinfo {author} {\bibfnamefont {D.~R.}\ \bibnamefont {Entem}}, \ and\
  \bibinfo {author} {\bibfnamefont {E.}~\bibnamefont {Ruiz~Arriola}},\ }\href
  {\doibase 10.1103/PhysRevC.81.024001} {\bibfield  {journal} {\bibinfo
  {journal} {Phys. Rev. C}\ }\textbf {\bibinfo {volume} {81}},\ \bibinfo
  {pages} {024001} (\bibinfo {year} {2010})},\ \Eprint
  {http://arxiv.org/abs/0910.3942} {arXiv:0910.3942 [nucl-th]} \BibitemShut
  {NoStop}%
\bibitem [{\citenamefont {Hu}\ \emph {et~al.}(2017)\citenamefont {Hu},
  \citenamefont {Zhang}, \citenamefont {Epelbaum}, \citenamefont
  {Mei\ss{}ner},\ and\ \citenamefont {Meng}}]{Hu:2016nkw}%
  \BibitemOpen
  \bibfield  {author} {\bibinfo {author} {\bibfnamefont {J.}~\bibnamefont
  {Hu}}, \bibinfo {author} {\bibfnamefont {Y.}~\bibnamefont {Zhang}}, \bibinfo
  {author} {\bibfnamefont {E.}~\bibnamefont {Epelbaum}}, \bibinfo {author}
  {\bibfnamefont {U.-G.}\ \bibnamefont {Mei\ss{}ner}}, \ and\ \bibinfo {author}
  {\bibfnamefont {J.}~\bibnamefont {Meng}},\ }\href {\doibase
  10.1103/PhysRevC.96.034307} {\bibfield  {journal} {\bibinfo  {journal} {Phys.
  Rev. C}\ }\textbf {\bibinfo {volume} {96}},\ \bibinfo {pages} {034307}
  (\bibinfo {year} {2017})},\ \Eprint {http://arxiv.org/abs/1612.05433}
  {arXiv:1612.05433 [nucl-th]} \BibitemShut {NoStop}%
\bibitem [{\citenamefont {Hebeler}\ \emph {et~al.}(2011)\citenamefont
  {Hebeler}, \citenamefont {Bogner}, \citenamefont {Furnstahl}, \citenamefont
  {Nogga},\ and\ \citenamefont {Schwenk}}]{Hebeler:2010xb}%
  \BibitemOpen
  \bibfield  {author} {\bibinfo {author} {\bibfnamefont {K.}~\bibnamefont
  {Hebeler}}, \bibinfo {author} {\bibfnamefont {S.~K.}\ \bibnamefont {Bogner}},
  \bibinfo {author} {\bibfnamefont {R.~J.}\ \bibnamefont {Furnstahl}}, \bibinfo
  {author} {\bibfnamefont {A.}~\bibnamefont {Nogga}}, \ and\ \bibinfo {author}
  {\bibfnamefont {A.}~\bibnamefont {Schwenk}},\ }\href {\doibase
  10.1103/PhysRevC.83.031301} {\bibfield  {journal} {\bibinfo  {journal} {Phys.
  Rev. C}\ }\textbf {\bibinfo {volume} {83}},\ \bibinfo {pages} {031301}
  (\bibinfo {year} {2011})},\ \Eprint {http://arxiv.org/abs/1012.3381}
  {arXiv:1012.3381 [nucl-th]} \BibitemShut {NoStop}%
\bibitem [{\citenamefont {Baardsen}\ \emph {et~al.}(2013)\citenamefont
  {Baardsen}, \citenamefont {Ekstr\"{o}m}, \citenamefont {Hagen},\ and\
  \citenamefont {Hjorth-Jensen}}]{cc_eos}%
  \BibitemOpen
  \bibfield  {author} {\bibinfo {author} {\bibfnamefont {G.}~\bibnamefont
  {Baardsen}}, \bibinfo {author} {\bibfnamefont {A.}~\bibnamefont
  {Ekstr\"{o}m}}, \bibinfo {author} {\bibfnamefont {G.}~\bibnamefont {Hagen}},
  \ and\ \bibinfo {author} {\bibfnamefont {M.}~\bibnamefont {Hjorth-Jensen}},\
  }\href {\doibase 10.1103/physrevc.88.054312} {\bibfield  {journal} {\bibinfo
  {journal} {Phys. Rev. C}\ }\textbf {\bibinfo {volume} {88}} (\bibinfo {year}
  {2013}),\ 10.1103/physrevc.88.054312}\BibitemShut {NoStop}%
\bibitem [{\citenamefont {Drischler}\ \emph {et~al.}(2019)\citenamefont
  {Drischler}, \citenamefont {Hebeler},\ and\ \citenamefont
  {Schwenk}}]{Drischler:2017wtt}%
  \BibitemOpen
  \bibfield  {author} {\bibinfo {author} {\bibfnamefont {C.}~\bibnamefont
  {Drischler}}, \bibinfo {author} {\bibfnamefont {K.}~\bibnamefont {Hebeler}},
  \ and\ \bibinfo {author} {\bibfnamefont {A.}~\bibnamefont {Schwenk}},\ }\href
  {\doibase 10.1103/PhysRevLett.122.042501} {\bibfield  {journal} {\bibinfo
  {journal} {Phys. Rev. Lett.}\ }\textbf {\bibinfo {volume} {122}},\ \bibinfo
  {pages} {042501} (\bibinfo {year} {2019})},\ \Eprint
  {http://arxiv.org/abs/1710.08220} {arXiv:1710.08220 [nucl-th]} \BibitemShut
  {NoStop}%
\bibitem [{\citenamefont {Sammarruca}\ and\ \citenamefont
  {Millerson}(2019)}]{Sammarruca:2019ncy}%
  \BibitemOpen
  \bibfield  {author} {\bibinfo {author} {\bibfnamefont {F.}~\bibnamefont
  {Sammarruca}}\ and\ \bibinfo {author} {\bibfnamefont {R.}~\bibnamefont
  {Millerson}},\ }\href {\doibase 10.3389/fphy.2019.00213} {\bibfield
  {journal} {\bibinfo  {journal} {Front. in Phys.}\ }\textbf {\bibinfo {volume}
  {7}},\ \bibinfo {pages} {213} (\bibinfo {year} {2019})}\BibitemShut {NoStop}%
\bibitem [{\citenamefont {Sammarruca}\ and\ \citenamefont
  {Millerson}(2020)}]{sam}%
  \BibitemOpen
  \bibfield  {author} {\bibinfo {author} {\bibfnamefont {F.}~\bibnamefont
  {Sammarruca}}\ and\ \bibinfo {author} {\bibfnamefont {R.}~\bibnamefont
  {Millerson}},\ }\href {\doibase 10.1103/PhysRevC.102.034313} {\bibfield
  {journal} {\bibinfo  {journal} {Phys. Rev. C}\ }\textbf {\bibinfo {volume}
  {102}},\ \bibinfo {pages} {034313} (\bibinfo {year} {2020})},\ \Eprint
  {http://arxiv.org/abs/2005.01958} {arXiv:2005.01958 [nucl-th]} \BibitemShut
  {NoStop}%
\bibitem [{\citenamefont {Drischler}\ \emph {et~al.}(2020)\citenamefont
  {Drischler}, \citenamefont {Melendez}, \citenamefont {Furnstahl},\ and\
  \citenamefont {Phillips}}]{dr_eos}%
  \BibitemOpen
  \bibfield  {author} {\bibinfo {author} {\bibfnamefont {C.}~\bibnamefont
  {Drischler}}, \bibinfo {author} {\bibfnamefont {J.~A.}\ \bibnamefont
  {Melendez}}, \bibinfo {author} {\bibfnamefont {R.~J.}\ \bibnamefont
  {Furnstahl}}, \ and\ \bibinfo {author} {\bibfnamefont {D.~R.}\ \bibnamefont
  {Phillips}},\ }\href {\doibase 10.1103/PhysRevC.102.054315} {\bibfield
  {journal} {\bibinfo  {journal} {Phys. Rev. C}\ }\textbf {\bibinfo {volume}
  {102}},\ \bibinfo {pages} {054315} (\bibinfo {year} {2020})},\ \Eprint
  {http://arxiv.org/abs/2004.07805} {arXiv:2004.07805 [nucl-th]} \BibitemShut
  {NoStop}%
\end{thebibliography}%
\bibliographystyle{apsrev4-1}

\end{document}